\documentclass[12pt,a4paper]{article}
\usepackage{ifthen} % for conditional statements
\newboolean{pdflatex}
\setboolean{pdflatex}{true} % False for eps figures 

\newboolean{articletitles}
\setboolean{articletitles}{true} % False removes titles in references

\newboolean{uprightparticles}
\setboolean{uprightparticles}{false} %True for upright particle symbols
\def\paperauthors{LHCb collaboration} % Leave as is for PAPER, CONF and FIGURE
\def\paperasciititle{Observation of the decay Bs -> Kz p pbar  and measurement of the Bds -> Kz p pbar branching fractions} % Set ASCII title here !! MAKE sure it's only ASCII characters !! 
\def\papertitle{Observation of the decay \BsToKzpp  and measurement of the \BdsToKzpp branching fractions} % Latex 
\def\paperkeywords{{High Energy Physics}, {LHCb}} % Comma separated list
\def\papercopyright{\the\year\ CERN for the benefit of the LHCb collaboration} % new since 9/Apr/2018
\def\paperlicence{CC BY 4.0 licence}
\def\paperlicenceurl{https://creativecommons.org/licenses/by/4.0/}

\newif\ifEnableSectionTOCLinks
\EnableSectionTOCLinksfalse % deactivated

\usepackage[top=1in, bottom=1.25in, left=1in, right=1in]{geometry}

\usepackage{microtype}
\usepackage{lineno}  % for line numbering during review
\usepackage{xspace} % To avoid problems with missing or double spaces after
\usepackage{caption} %these three command get the figure and table captions automatically small

\usepackage{graphicx}  % to include figures (can also use other packages)
\usepackage{color}
\usepackage{colortbl}
\graphicspath{{./figs/}} % Make Latex search fig subdir for figures
\usepackage{amsmath} % Adds a large collection of math symbols
\usepackage{amssymb}
\usepackage{amsfonts}
\usepackage{upgreek} % Adds in support for greek letters in roman typeset

\newcommand*\patchAmsMathEnvironmentForLineno[1]{%
\expandafter\let\csname old#1\expandafter\endcsname\csname #1\endcsname
\expandafter\let\csname oldend#1\expandafter\endcsname\csname
end#1\endcsname
 \renewenvironment{#1}%
   {\linenomath\csname old#1\endcsname}%
   {\csname oldend#1\endcsname\endlinenomath}%
}
\newcommand*\patchBothAmsMathEnvironmentsForLineno[1]{%
  \patchAmsMathEnvironmentForLineno{#1}%
  \patchAmsMathEnvironmentForLineno{#1*}%
}
\AtBeginDocument{%
\patchBothAmsMathEnvironmentsForLineno{equation}%
\patchBothAmsMathEnvironmentsForLineno{align}%
\patchBothAmsMathEnvironmentsForLineno{flalign}%
\patchBothAmsMathEnvironmentsForLineno{alignat}%
\patchBothAmsMathEnvironmentsForLineno{gather}%
\patchBothAmsMathEnvironmentsForLineno{multline}%
\patchBothAmsMathEnvironmentsForLineno{eqnarray}%
}

\usepackage[pdftex,
            pdfauthor={\paperauthors},
            pdftitle={\paperasciititle},
            pdfkeywords={\paperkeywords}]{hyperref}
\usepackage{hyperxmp}
\hypersetup{
    pdfcopyright={Copyright (C) \papercopyright},
    pdflicenseurl={\paperlicenceurl}
}
\usepackage[colorinlistoftodos,textsize=scriptsize]{todonotes}
\usepackage[bottom,flushmargin,hang,multiple]{footmisc}

\usepackage[all]{hypcap} % Internal hyperlinks to floats.

\usepackage{xspace} 
\usepackage{upgreek}

\def\lhcb   {\mbox{LHCb}\xspace}

\def\babar  {\mbox{BaBar}\xspace}
\def\belle  {\mbox{Belle}\xspace}

\def\MagUp {\mbox{\em Mag\kern -0.05em Up}\xspace}

\ifthenelse{\boolean{uprightparticles}}%
{

 \def\Peta        {\ensuremath{\upeta}\xspace}

 \def\Ppi         {\ensuremath{\uppi}\xspace}

 \def\Ppsi        {\ensuremath{\uppsi}\xspace}

 \def\PDelta      {\ensuremath{\Delta}\xspace}                 
 \def\PXi         {\ensuremath{\Xi}\xspace}                 
 \def\PLambda     {\ensuremath{\Lambda}\xspace}                 
 \def\PSigma      {\ensuremath{\Sigma}\xspace}                 
 \def\POmega      {\ensuremath{\Omega}\xspace}                 
 \def\PUpsilon    {\ensuremath{\Upsilon}\xspace}
 \let\oldPi\Pi
 \def\PPi         {\ensuremath{\oldPi}\xspace}

 \def\PB      {\ensuremath{\mathrm{B}}\xspace}                 
 \def\PD      {\ensuremath{\mathrm{D}}\xspace}                 
 \def\PJ      {\ensuremath{\mathrm{J}}\xspace}                 
 \def\PK      {\ensuremath{\mathrm{K}}\xspace}                 
 \def\Pb      {\ensuremath{\mathrm{b}}\xspace}                 
 \def\Pc      {\ensuremath{\mathrm{c}}\xspace}                 
 \def\Pd      {\ensuremath{\mathrm{d}}\xspace}                 

 \def\Pp      {\ensuremath{\mathrm{p}}\xspace}                 
 \def\Pq      {\ensuremath{\mathrm{q}}\xspace}                 
 \def\Ps      {\ensuremath{\mathrm{s}}\xspace}                 
                  
 \def\Pu      {\ensuremath{\mathrm{u}}\xspace}                 
 \def\thebaroffset{0.0em}
}
{

 \def\Peta        {\ensuremath{\eta}\xspace}

 \def\Ppi         {\ensuremath{\pi}\xspace}

 \def\Ppsi        {\ensuremath{\psi}\xspace}                 
                  
 \mathchardef\PDelta="7101
 \mathchardef\PXi="7104
 \mathchardef\PLambda="7103
 \mathchardef\PSigma="7106
 \mathchardef\POmega="710A
 \mathchardef\PUpsilon="7107
 \mathchardef\PPi="7105
 \def\PB      {\ensuremath{B}\xspace}                 
 \def\PD      {\ensuremath{D}\xspace}                 
 \def\PJ      {\ensuremath{J}\xspace}                 
 \def\PK      {\ensuremath{K}\xspace}                 
 \def\Pb      {\ensuremath{b}\xspace}                 
 \def\Pc      {\ensuremath{c}\xspace}                 
 \def\Pd      {\ensuremath{d}\xspace}                 

 \def\Pp      {\ensuremath{p}\xspace}                 
 \def\Pq      {\ensuremath{q}\xspace}                 
 \def\Ps      {\ensuremath{s}\xspace}                 
                  
 \def\Pu      {\ensuremath{u}\xspace}                 
 \def\thebaroffset{0.18em}
}
\newcommand{\offsetoverline}[2][\thebaroffset]{\kern #1\overline{\kern -#1 #2}}%

\makeatletter
\ifcase \@ptsize \relax% 10pt
  \newcommand{\miniscule}{\@setfontsize\miniscule{4}{5}}% \tiny: 5/6
\or% 11pt
  \newcommand{\miniscule}{\@setfontsize\miniscule{5}{6}}% \tiny: 6/7
\or% 12pt
  \newcommand{\miniscule}{\@setfontsize\miniscule{5}{6}}% \tiny: 6/7
\fi
\makeatother

\DeclareRobustCommand{\optbar}[1]{\shortstack{{\miniscule (\rule[.5ex]{1.25em}{.18mm})}
  \\ [-.7ex] $#1$}}

\def\quark     {{\ensuremath{\Pq}}\xspace}
\def\quarkbar  {{\ensuremath{\overline \quark}}\xspace}
\def\qqbar     {{\ensuremath{\quark\quarkbar}}\xspace}
\def\uquark    {{\ensuremath{\Pu}}\xspace}

\def\dquark    {{\ensuremath{\Pd}}\xspace}

\def\squark    {{\ensuremath{\Ps}}\xspace}

\def\cquark    {{\ensuremath{\Pc}}\xspace}

\def\bquark    {{\ensuremath{\Pb}}\xspace}

\def\pion   {{\ensuremath{\Ppi}}\xspace}
\def\piz    {{\ensuremath{\pion^0}}\xspace}
\def\pip    {{\ensuremath{\pion^+}}\xspace}
\def\pim    {{\ensuremath{\pion^-}}\xspace}

\def\pimp   {{\ensuremath{\pion^\mp}}\xspace}

\def\kaon    {{\ensuremath{\PK}}\xspace}
\def\KorKbar {\kern \thebaroffset\optbar{\kern -\thebaroffset \PK}{}\xspace}
\def\Kz      {{\ensuremath{\kaon^0}}\xspace}
\def\Kzu      {{\ensuremath{\kaon^{0,+}}}\xspace}

\def\Kp      {{\ensuremath{\kaon^+}}\xspace}
\def\Km      {{\ensuremath{\kaon^-}}\xspace}
\def\Kpm     {{\ensuremath{\kaon^\pm}}\xspace}

\def\KS      {{\ensuremath{\kaon^0_{\mathrm{S}}}}\xspace}

\def\D       {{\ensuremath{\PD}}\xspace}

\def\DorDbar {\kern \thebaroffset\optbar{\kern -\thebaroffset \PD}\xspace}
\def\Dz      {{\ensuremath{\D^0}}\xspace}

\def\Dp      {{\ensuremath{\D^+}}\xspace}
\def\Dm      {{\ensuremath{\D^-}}\xspace}

\def\DpDm    {\ensuremath{\Dp {\kern -0.16em \Dm}}\xspace}

\def\Dstarp  {{\ensuremath{\D^{*+}}}\xspace}

\def\B       {{\ensuremath{\PB}}\xspace}

\def\BorBbar {\kern \thebaroffset\optbar{\kern -\thebaroffset \PB}\xspace}
\def\Bz      {{\ensuremath{\B^0}}\xspace}

\def\Bd      {{\ensuremath{\B^0}}\xspace}

\def\BdorBdbar {\kern \thebaroffset\optbar{\kern -\thebaroffset \Bd}\xspace}
\def\Bu      {{\ensuremath{\B^+}}\xspace}
\def\Bdu      {{\ensuremath{\B^{0,+}}}\xspace}

\def\Bs      {{\ensuremath{\B^0_\squark}}\xspace}

\def\BsorBsbar {\kern \thebaroffset\optbar{\kern -\thebaroffset \Bs}\xspace}

\def\Bds     {{\ensuremath{\B_{(\squark)}^0}}\xspace}

\def\jpsi     {{\ensuremath{{\PJ\mskip -3mu/\mskip -2mu\Ppsi}}}\xspace}
\def\psitwos  {{\ensuremath{\Ppsi{(2S)}}}\xspace}

\def\etac     {{\ensuremath{\Peta_\cquark}}\xspace}

\def\Y#1S{\ensuremath{\PUpsilon{(#1S)}}\xspace}

\def\proton      {{\ensuremath{\Pp}}\xspace}
\def\antiproton  {{\ensuremath{\overline \proton}}\xspace}

\def\Lz          {{\ensuremath{\PLambda}}\xspace}

\def\LorLbar     {\kern \thebaroffset\optbar{\kern -\thebaroffset \PLambda}\xspace}

\def\Lc          {{\ensuremath{\Lz^+_\cquark}}\xspace}

\def\BF         {{\ensuremath{\mathcal{B}}}\xspace}
\def\BR         {\BF}

\newcommand{\decay}[2]{\mbox{\ensuremath{#1\!\to #2}}\xspace} 

\def\to                 {\ensuremath{\rightarrow}\xspace}

\def\CP                {{\ensuremath{C\!P}}\xspace}

\def\AT#1     {\ensuremath{A_{\mathrm{T}}^{#1}}\xspace}           % 2

\def\C#1      {\ensuremath{\mathcal{C}_{#1}}\xspace}                       % 9
\def\Cp#1     {\ensuremath{\mathcal{C}_{#1}^{'}}\xspace}                    % 7
\def\Ceff#1   {\ensuremath{\mathcal{C}_{#1}^{\mathrm{(eff)}}}\xspace}        % 9  
\def\Cpeff#1  {\ensuremath{\mathcal{C}_{#1}^{'\mathrm{(eff)}}}\xspace}       % 7
\def\Ope#1    {\ensuremath{\mathcal{O}_{#1}}\xspace}                       % 2
\def\Opep#1   {\ensuremath{\mathcal{O}_{#1}^{'}}\xspace}                    % 7

\newcommand{\nospaceunit}[1]{\ensuremath{\text{#1}}}       
\newcommand{\aunit}[1]{\ensuremath{\text{\,#1}}}       
\newcommand{\tev}{\aunit{Te\kern -0.1em V}\xspace}
\newcommand{\gev}{\aunit{Ge\kern -0.1em V}\xspace}
\newcommand{\mev}{\aunit{Me\kern -0.1em V}\xspace}
\newcommand{\kev}{\aunit{ke\kern -0.1em V}\xspace}
\newcommand{\ev}{\aunit{e\kern -0.1em V}\xspace}
 
\newcommand{\mevc}{\ensuremath{\aunit{Me\kern -0.1em V\!/}c}\xspace}
\newcommand{\gevc}{\ensuremath{\aunit{Ge\kern -0.1em V\!/}c}\xspace}
\newcommand{\mevcc}{\ensuremath{\aunit{Me\kern -0.1em V\!/}c^2}\xspace}
\newcommand{\gevcc}{\ensuremath{\aunit{Ge\kern -0.1em V\!/}c^2}\xspace}
\def\mm   {\aunit{mm}\xspace}

\def\mum  {\ensuremath{\,\upmu\nospaceunit{m}}\xspace}

\def\fb   {\ensuremath{\aunit{fb}}\xspace}
\def\invfb   {\ensuremath{\fb^{-1}}\xspace}

\newcommand{\chisq}{\ensuremath{\chi^2}\xspace}

\def\gsim{{~\raise.15em\hbox{$>$}\kern-.85em
          \lower.35em\hbox{$\sim$}~}\xspace}
\def\lsim{{~\raise.15em\hbox{$<$}\kern-.85em
          \lower.35em\hbox{$\sim$}~}\xspace}

\def\sPlot{\mbox{\em sPlot}\xspace}

\def\pt         {\ensuremath{p_{\mathrm{T}}}\xspace}

\def\ptot       {\ensuremath{p}\xspace}

\def\evtgen     {\mbox{\textsc{EvtGen}}\xspace}

\def\geant      {\mbox{\textsc{Geant4}}\xspace}

\def\photos     {\mbox{\textsc{Photos}}\xspace}

\def\pythia     {\mbox{\textsc{Pythia}}\xspace}

\def\tell1  {TELL1\xspace}
\def\ukl1   {UKL1\xspace}

\newcommand{\lhcborcid}[1]{\href{https://orcid.org/#1}{\hspace*{0.1em}\raisebox{-0.45ex}{\includegraphics[width=1em]{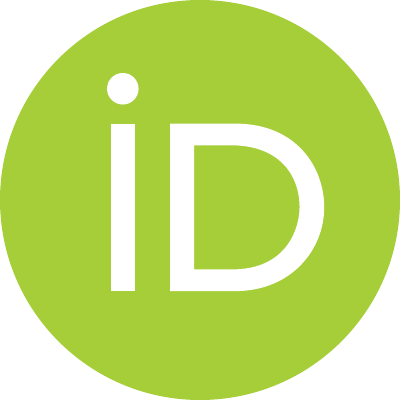}}}}

\def\BdToKSpp     {\decay{\Bd}{\KS\proton\antiproton}}
\def\BsToKSpp     {\decay{\Bs}{\KS\proton\antiproton}}
\def\BdsToKSpp    {\decay{\Bds}{\KS\proton\antiproton}}
\def\BdToKSpipi   {\decay{\Bd}{\KS\pip\pim}}

\def\BdsToKShh    {\decay{\Bds}{\KS h^+ h^-}}
\def\BdToKShh    {\decay{\Bd}{\KS h^+ h^-}}

\def\BdsToKSKpi   {\decay{\Bds}{\KS\Kpm\pimp}}
\def\BdsToKSKpi   {\decay{\Bds}{\KS\Kpm\pimp}}

\def\KSTopipi     {\decay{\KS}{\pip\pim}}
\def\DstarToDpiToKpi     {\decay{\Dstarp}{\decay{\Dz(}{\Km\pip)}}\pip}
\def\BdTojpsiKpi     {\decay{\Bd}{\jpsi\Kp\pim}}

\def\KShh   {{\ensuremath{\KS h^+ h^-}}\xspace}

\def\ppbar   {{\ensuremath{\proton\antiproton}}\xspace}
\def\KSpp   {{\ensuremath{\KS\ppbar}}\xspace}
\def\KSpipi   {{\ensuremath{\KS\pip\pim}}\xspace}

\def\BdToKzpipi   {\decay{\Bd}{\Kz\pip\pim}}
\def\BdToKzpp   {\decay{\Bd}{\Kz\proton\antiproton}}
\def\BsToKzpp   {\decay{\Bs}{\Kz\proton\antiproton}}
\def\BdsToKzpp    {\decay{\Bds}{\Kz\proton\antiproton}}
\def\LzToppi    {\decay{\Lz}{\proton\pim}}

\def\BuToKpp   {\decay{\Bu}{\Kp\proton\antiproton}}

\def\BduTopiK   {\decay{\Bdu}{\piz\Kzu}}
\def\BduToJpiK   {\decay{\Bdu}{\jpsi\Kzu}}

\def\BdKSppResult           {(5.67 \pm 0.17 \pm 0.28) \times 10^{-2}}

\def\BsKSppResult           {(1.84 \pm 0.34 \pm 0.18 \pm 0.04) \times 10^{-2}}
\def\BsKSppResultBd           {(3.25 \pm 0.61 \pm 0.36 \pm 0.07) \times 10^{-1}}

\def\BdKzpipiResultBF          {(4.97 \pm 0.18) \times 10^{-5}}
\def\BdKzppResultBF           {(2.82 \pm 0.08 \pm 0.12 \pm 0.10) \times 10^{-6}}
\def\BsKzppResultBF           {(9.14 \pm 1.69 \pm 0.90 \pm 0.33 \pm 0.20) \times 10^{-7}}
\def\BdKzppResultBFPDG           {(2.66 \pm 0.32) \times 10^{-6}}

\def\sPlot{\mbox{\em sPlot}\xspace}

\def\BsL      {{\ensuremath{\B^0_{\squark\mathrm{L}}}}\xspace}
\def\BsH      {{\ensuremath{\B^0_{\squark\mathrm{H}}}}\xspace} % Add in the predefined LHCb symbols

\hypersetup{
  colorlinks   = true, %Colours links instead of ugly boxes
  urlcolor     = blue, %Colour for external hyperlinks
  linkcolor    = blue, %Colour of internal links
  citecolor    = red   %Colour of citations
}

\ifEnableSectionTOCLinks
    \usepackage[explicit]{titlesec} % to change headings
    
    \let\oldcontentsline\contentsline
    \renewcommand

    \titleformat{\section}{\normalfont\Large\bf}{\hyperlink{tocsection.\thesection}{{\thesection} \parbox[t]{\dimexpr\textwidth-1pc}{#1}}}{1pc}{}

    \titleformat{\subsection}{\normalfont\bf}{\hyperlink{tocsubsection.\thesubsection}{{\thesubsection} \parbox[t]{\dimexpr\textwidth-1pc}{#1}}}{1pc}{}

    \titleformat{name=\section,numberless}[display]{}{}{0pt}{\normalfont\Huge\bfseries #1}
\fi

\usepackage{cite} % Allows for ranges in citations
\usepackage{mciteplus}
\usepackage{longtable} % only for template; not usually to be used in PAPERs
\usepackage{multirow}
\usepackage{float}

\begin{document}

%%%%%%%%%%%%%%%%%%%%%%%%%
%%%%% Title     %%%%%%%%%
%%%%%%%%%%%%%%%%%%%%%%%%%
\renewcommand{\thefootnote}{\fnsymbol{footnote}}
\setcounter{footnote}{1}
% %%%%%%% CHOOSE TITLE PAGE--------
%\onecolumn
%\input{title-LHCb-INT}
%\input{title-LHCb-ANA}
%\input{title-LHCb-CONF}
%\input{title-LHCb-FIGURE}
% $Id: title-LHCb-PAPER.tex 111203 2017-08-08 15:28:40Z pkoppenb $
% ===============================================================================
% Purpose: LHCb-PAPER journal paper title page template
% Author: 
% Created on: 2010-09-25
% ===============================================================================

%%%%%%%%%%%%%%%%%%%%%%%%%
%%%%%  TITLE PAGE  %%%%%%
%%%%%%%%%%%%%%%%%%%%%%%%%
\begin{titlepage}
\pagenumbering{roman}

% Header ---------------------------------------------------
\vspace*{-1.5cm}
\centerline{\large EUROPEAN ORGANIZATION FOR NUCLEAR RESEARCH (CERN)}
\vspace*{1.5cm}
\noindent
\begin{tabular*}{\linewidth}{lc@{\extracolsep{\fill}}r@{\extracolsep{0pt}}}
\ifthenelse{\boolean{pdflatex}}% Logo format choice
{\vspace*{-1.5cm}\mbox{\!\!\!\includegraphics[width=.14\textwidth]{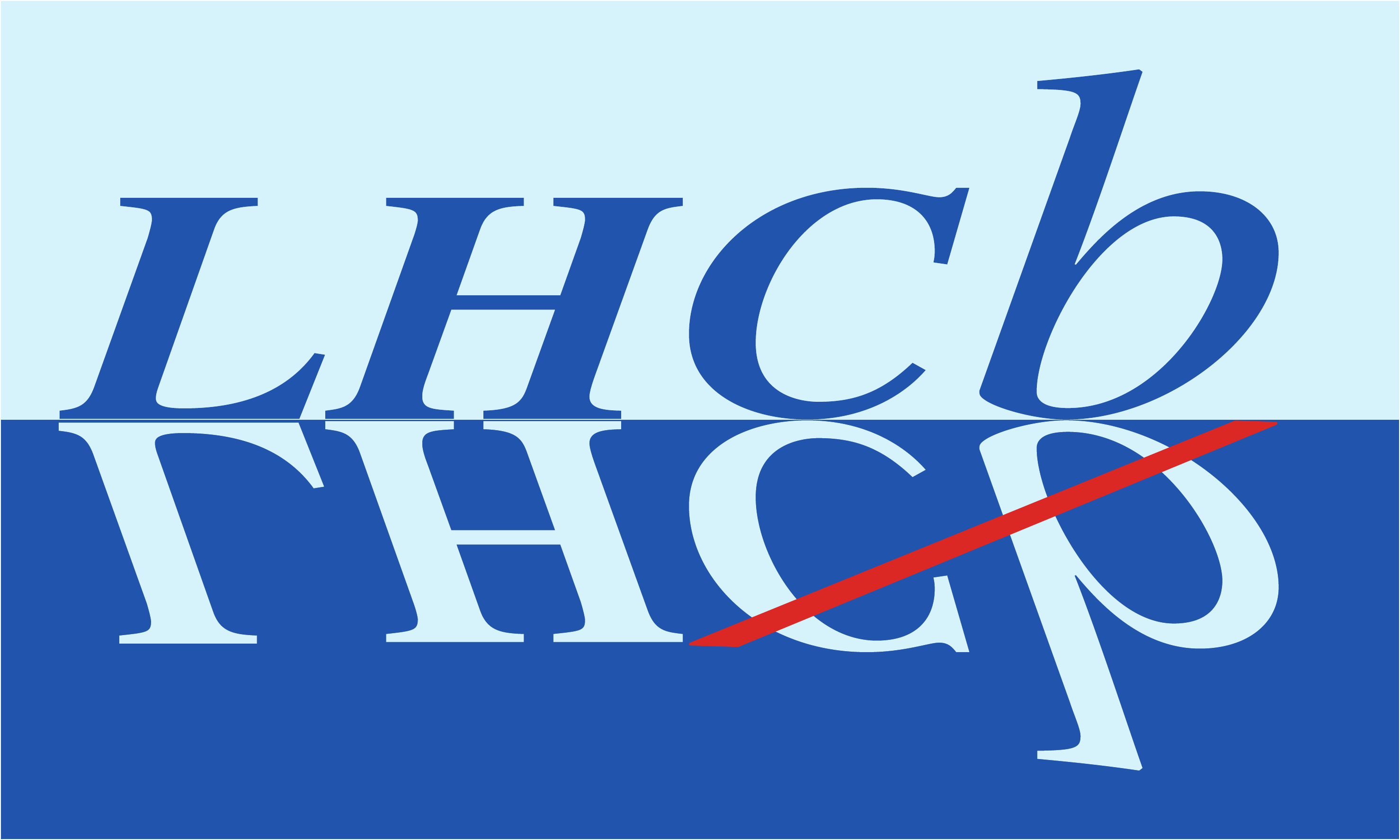}} & &}%
{\vspace*{-1.2cm}\mbox{\!\!\!\includegraphics[width=.12\textwidth]{figs/lhcb-logo.eps}} & &}%
\\
 & & CERN-EP-2025-085 \\  % ID 
 & & LHCb-PAPER-2025-001 \\  % ID 
 & & July 10, 2025 \\ % Date - Can also hardwire e.g.
\end{tabular*}

\vspace*{2.0cm}

% Title --------------------------------------------------
{\normalfont\bfseries\boldmath\huge
\begin{center}
% DO NOT EDIT HERE. Instead edit macro in main.tex to keep metadata correct
  \papertitle 
\end{center}
}

\vspace*{1.5cm}

% Authors -------------------------------------------------
\begin{center}
%In the footnote, replace 'paper' by 'Letter' in case of submission to PRL or PLB 
% Edit macro in main.tex to keep metadata correct
\paperauthors\footnote{Authors are listed at the end of this paper.}
\end{center}

\vspace{\fill}

% Abstract -----------------------------------------------
\begin{abstract}
  \noindent
A study of the charmless baryonic decays \BdsToKzpp is presented, where \Bds denotes either a \Bd or 
a \Bs meson. The analysis is based on proton-proton collision data collected by the \lhcb experiment at centre-of-mass energies of 7, 8, and 13\tev, corresponding to an integrated luminosity of 9\invfb. The decay \BsToKzpp is observed for the first time, with a measured branching fraction of \mbox{$\BsKzppResultBF$} and a significance of 5.6$\sigma$. The uncertainties respectively account for statistical and systematic contributions, the precision of the branching fraction of the normalisation channel \BdToKzpipi and the fragmentation fraction ratio ${f_s}/{f_d}$. The branching fraction determined for \BdToKzpp is \mbox{$\BdKzppResultBF$},  which is the most precise measurement to date.
\end{abstract}

\vspace*{2.0cm}

\begin{center}
  Published in JHEP 07 (2025) 121
\end{center}

\vspace{\fill}

{\footnotesize 
% Edit macro in main.tex to keep metadata correct
\centerline{\copyright~\papercopyright. \href{\paperlicenceurl}{\paperlicence}.}}
\vspace*{2mm}

\end{titlepage}

%%%%%%%%%%%%%%%%%%%%%%%%%%%%%%%%
%%%%%  EOD OF TITLE PAGE  %%%%%%
%%%%%%%%%%%%%%%%%%%%%%%%%%%%%%%%

%  empty page follows the title page ----
\newpage
\setcounter{page}{2}
\mbox{~}

\cleardoublepage

%\twocolumn
% %%%%%%%%%%%%% ---------

\renewcommand{\thefootnote}{\arabic{footnote}}
\setcounter{footnote}{0}

%%%%%%%%%%%%%%%%%%%%%%%%%%%%%%%%
%%%%%  Table of Content   %%%%%%
%%%%%%%%%%%%%%%%%%%%%%%%%%%%%%%%
%%%% Uncomment if desired
%\tableofcontents

\cleardoublepage

%%%%%%%%%%%%%%%%%%%%%%%%%
%%%%% Main text %%%%%%%%%
%%%%%%%%%%%%%%%%%%%%%%%%%

\pagestyle{plain} % restore page numbers for the main text
\setcounter{page}{1}
\pagenumbering{arabic}

%% Uncomment during review phase. 
%% Comment before a final submission.
%% \linenumbers

%% This is the main body
%% It is useful to have a single file so comments are not missed in overleaf.
%%%%%%%%%%%%%%%%%%%%%%%%%%%%%%%%%%%%%%%%%%%%%%%%%%%%%%%%%%%%%%%%%%%%%%%%%%%%%%%%%%%%%%%%%%%%%%%%%%
\section{Introduction }
\label{sec:Introduction}
The \BdsToKzpp decays involve contributions from both QCD penguin and tree-level amplitudes. The QCD penguin contributions proceed via the \decay{\bquark}{\qqbar\squark} and \decay{\bquark}{\qqbar\dquark} quark transitions (where $q=u, d$), while the tree-level \decay{\bquark}{\uquark} contributions are CKM suppressed~\cite{Cabibbo:1963yz}. These processes provide a valuable probe into the dynamics of charmless \B meson decays. The \ppbar mass distribution that peaks near threshold in these decays is a key feature indicating the involvement of nontrivial intermediate states that influence the decay dynamics. The observed decay rate hierarchy, where the \BdToKzpp rate is significantly lower than that of the \BuToKpp mode, contrasts with other decay modes such as \BduTopiK and \BduToJpiK~\cite{Rosner_2003, Wang_2004, Yuan_2005, Suzuki_2006, LAPORTA_2007, Belle:2008zkc}. 

Measurements of branching fractions and \CP asymmetries are essential for testing and refining theoretical model~\cite{Chua_2002,Geng_2006}. The relative branching fractions of \BdToKzpp and \BsToKzpp decays provide precise tests of flavour symmetries, as their quark-level diagrams are related not only through the interchange of \dquark and \squark spectator quarks but also through modifications to the weak decay process to ensure the correct final-state quark content. Additionally, these decays serve as key observables for studying \CP violation in loop-dominated processes~\cite{LHCb-PAPER-2018-051}. Time-dependent Dalitz analyses can measure \CP-violating parameters, which may be sensitive to potential physics beyond the Standard Model~\cite{Garmash_2007, Aubert_2009}.

In this paper, a search for the \BsToKzpp decay mode is performed, and the branching fractions of \BdsToKzpp decays are determined by employing the well-known \BdToKzpipi decays for normalisation~\cite{PDG2024}. In both decays, the \Kz is reconstructed through the two-pion decay of its short-lived mass eigenstate (\KSTopipi).
This analysis uses the full Run~1 (2011--2012) and Run~2 (2015--2018) LHCb data samples of proton-proton (\proton\proton) collisions at centre-of-mass energies of 7, 8, and 13\tev, corresponding to an integrated luminosity of about 9\invfb. To improve statistical significance, the data sample is divided into four subsamples (2011, 2012, 2015--2016, and 2017--2018) based on variations in peak shapes, which arise from differences in the centre-of-mass energies and trigger configurations during data collection.

While the measured \proton\antiproton and \Kz\proton mass regions span the full kinematic range, contributions from charmonium (\etac, \jpsi, \psitwos) and open-charm (\Lc) decays are explicitly excluded from the respective regions. A dedicated efficiency correction accounts for the phase-space contribution in the excluded regions, ensuring that the aforementioned mass vetoes have a negligible impact on the branching fraction measurement. The \BdToKzpp branching fractions have been measured by the \babar and \belle collaborations~\cite{BaBar:2007esv, Belle:2008zkc}, with both results in good agreement. These previous measurements also cover the full \proton\antiproton kinematic range, allowing a direct comparison.

%%%%%%%%%%%%%%%%%%%%%%%%%%%%%%%%%%%%%%%%%%%%%%%%%%%%%%%%%%%%%%%%%%%%%%%%%%%%%%%%%%%%%%%%%%%%%%%%%%
\section{Detector and simulation}
\label{sec:Detector:simulation}
The \lhcb detector~\cite{LHCb-DP-2008-001,LHCb-DP-2014-002} is a single-arm forward spectrometer covering the pseudorapidity range $2<\eta <5$, designed for the study of particles containing \bquark or \cquark quarks. 
The detector includes a high-precision tracking system consisting of a silicon-strip vertex detector surrounding the $pp$ interaction region~\cite{LHCb-DP-2014-001}, 
a large-area silicon-strip detector located upstream of a dipole magnet with a bending power of about $4{\mathrm{\,T\,m}}$, and three stations of silicon-strip detectors and straw drift tubes~\cite{LHCb-DP-2013-003} placed downstream of the magnet.
The tracking system provides a measurement of momentum, \ptot, of charged particles with a relative uncertainty that varies from 0.5\% at low momentum to 1.0\% at 200\gevc. 
The minimum distance of a track to a primary vertex (PV), the impact parameter (IP), is measured with a resolution of $(15+29/\pt)\mum$, 
where \pt is the component of the momentum transverse to the beam, in\,\gevc. 
Different types of charged hadrons are distinguished using information from two ring-imaging Cherenkov detectors~\cite{LHCb-DP-2012-003}. 
Photons, electrons and hadrons are identified by a calorimeter system consisting of scintillating-pad and preshower detectors, an electromagnetic calorimeter and a hadronic calorimeter. 
Muons are identified by a system composed of alternating layers of iron and multiwire proportional chambers~\cite{LHCb-DP-2012-002}.

Simulation is used to model the effects of detector acceptance and selection requirements, as well as to study backgrounds from other $b$-hadron decays. Both the signal and normalisation decays are simulated using a phase-space model.
In the simulation, $pp$ collisions are generated using \pythia with a specific \lhcb configuration~\cite{LHCb-PROC-2010-056}. 
Decays of unstable particles are described by \evtgen~\cite{Lange:2001uf}, in which final-state radiation is generated using \photos~\cite{Golonka:2005pn}. 
The interaction of the generated particles with the detector, and its response, are implemented using the \geant
toolkit~\cite{Allison:2006ve, Agostinelli:2002hh} as described in Ref.~\cite{LHCb-PROC-2011-006}.

The online event selection is performed by a trigger~\cite{LHCb-DP-2012-004}, which consists of a hardware stage, based on information from the calorimeter and muon systems, followed by a software stage, which applies a full event reconstruction. 
At the hardware trigger level, events are required to have a muon with high \pt or a hadron, photon or electron with high transverse energy in the calorimeters. 
The software trigger requires a two-, three- or four-track secondary vertex with a significant primary \proton\proton interaction vertices with PVs. 
At least one charged particle must have high transverse momentum and be inconsistent with originating from any reconstructed PV. 
A multivariate algorithm~\cite{BBDT,LHCb-PROC-2015-018} is used for the identification of secondary vertices consistent with the decay of a \bquark hadron. It is required that the software trigger decision must have been caused by tracks from the decay of the signal \B candidate.

%%%%%%%%%%%%%%%%%%%%%%%%%%%%%%%%%%%%%%%%%%%%%%%%%%%%%%%%%%%%%%%%%%%%%%%%%%%%%%%%%%%%%%%%%%%%%%%%%%
\section{Event selection}
\label{sec:Trigger:event：selection}
To suppress background, events satisfying the hardware and software trigger requirements undergo an additional two-stage offline selection: a loose preselection followed by a multivariate selection. The loose preselection, which primarily relies on topological features to distinguish signal from background, includes criteria such as the displacement of the \B decay products from the PV.

Reconstructed \decay{\KS}{\pip\pim} decays are classified into two categories: the `long' category, which includes \KS mesons that decay at shorter flight distances, allowing pion tracks to be reconstructed in the vertex detector, and the `downstream' category, which comprises \KS mesons that decay further downstream, where the pion track segments cannot be formed in the vertex detector.
The long category has better mass, momentum and vertex resolution than the downstream category.
The \KS candidates are formed by combining two charged pions associated with long (downstream) tracks, with momenta greater than 2\gevc (6\gevc), and inconsistent with originating from a PV; the dipion vertex fit must be of good quality, and the resulting invariant mass must lie within $\pm20\mevc$ ($\pm30\mevc$) of the known \KS mass~\cite{PDG2024}. The \KS vertex must be significantly displaced from its associated PV, which is defined using the difference in the vertex-fit \chisq of a given PV reconstructed with and without the \KS candidate. 

The \B candidates are formed by combining a \KS candidate with two oppositely charged hadrons, either \ppbar or \pip\pim. In the former case, to ensure good particle identification (PID), the proton momentum is required to be within $8 < p < 100\gevc$ for Run~1 and $11 < p < 100\gevc$ for Run~2~\cite{LHCb-PAPER-2022-003}.
The scalar sum of the transverse momenta of the \KS meson and two proton candidates must be greater than $3.0 \gevc\ (4.2 \gevc)$, for long (downstream) candidates, and at least two of these three decay products must have $\pt > 0.8\gevc$. The reconstructed \B candidate vertex is required to be of good quality, significantly separated from any PV, and isolated from other tracks.
The \B candidates should also be consistent with originating from a PV.
Finally, the \KS decay vertex is required to be at least 30\mm downstream of the \B vertex along the beam direction. The PID requirements are applied to ensure each final-state particle is consistent with its hypothesis.

A boosted decision tree (BDT) classifier~\cite{Breiman,AdaBoost}, implemented in the TMVA toolkit~\cite{Hocker:2007ht}, is used to separate signal from background. The classifier is trained using variables related to the \B candidate, such as \pt, pseudorapidity, decay time, the vertex-fit quality, and the angle between its reconstructed momentum and flight direction. It also considers vertex properties such as the consistency with the PV, the separation between the \B and \KS vertices along the beam direction, and the impact parameter significance and \pt of the final-state particles.

As the decay topologies of \BdsToKSpp and \BdToKSpipi are nearly identical, 
%to enhance sensitivity to the \BsToKSpp channel, 
only \BdToKSpp decay samples are used to train the classifier. The signal training sample consists of simulated \BdToKSpp decays, while the background training sample consists of data from the upper-mass sideband region, defined as  
$5450 < m(\KSpp) < 5800\mevcc$.

The \BdsToKSpp and \BdToKSpipi candidates are categorised by \KS reconstruction type (long and downstream) and data-taking period (2011 and 2012 share a common selection). Selections are optimised separately for the \BdToKSpp branching fraction measurement and the \BsToKSpp search, with the BDT requirement optimised independently for each subsample. For the previously established \BdToKShh decays, with h=$\pi$ or \proton, the BDT selection is optimised for signal significance, defined as $N_{\rm sig}/\sqrt{N_{\rm sig} + N_{\rm bkg}}$, where $N_{\rm sig}$ and $N_{\rm bkg}$ represent the numbers of expected signal and background candidates within the signal $5234 < m(\KShh) < 5324\mevcc$.
The value of $N_{\rm sig}$ is estimated using the signal yield extracted from fits to data selected with a looser BDT selection requirement and scaling it by the efficiency of the tighter BDT selection, as determined from simulation. The value of $N_{\rm bkg}$ is calculated by fitting the upper mass sideband region and extrapolating into the signal region.
In contrast, since the \BsToKSpp decay mode has not yet been observed, the BDT selection is optimised to enhance its sensitivity, based on the figure-of-merit 
$\varepsilon_{\rm sig}/(\alpha/2 + \sqrt{N_{\rm bkg}})$~\cite{Punzi:2003bu}, 
%%%%%%%%%%%%%%%%%%%%%%%%%%%%%%%%%%%%%%%%%%%%%%%%%%%%%%%%%%%%%%%%%%
%%%%%%%%%%%%%%%%%%%%%%%%%%%%%%%%%%%%%%%%%%%%%%%%%%%%%%%%%%%%%%%%%%
where $\varepsilon_{\rm sig}$ is the signal efficiency, estimated from simulation, and $\alpha=5$ is the desired significance for the signal channel.

After the selection process, the background includes \B decays with similar or identical final states that proceed through intermediate open-charm or charmonium states. Dedicated invariant-mass vetoes around the known mass values~\cite{PDG2024} are applied to reject these backgrounds. 
Specifically, the veto windows are set to $\pm30 \mevcc$ for the \Lc baryon, $\pm45 \mevcc$ for the \etac state, $\pm25 \mevcc$ for the \jpsi state, and $\pm35 \mevcc$ for the \psitwos state. These ranges typically correspond to $\pm 3 \sigma$, where $\sigma$ is the mass resolution of the considered resonance.
The effect of the vetoes is determined using simulations, which account for the exclusion of the \ppbar and \KS\proton regions. 

\section{Signal extraction}
\label{sec:Fit:model}  
An unbinned maximum-likelihood fit to the invariant mass is performed for each of the eight subsamples; the invariant masses are determined by constraining the \KS mass to its known value~\cite{PDG2024}. The 2011 and 2012 data share the candidate selection but have separate mass fits and efficiency calculations due to variations in peak shapes, within the mass regions $5000 < m(\KSpp) < 5600\mevcc$ for \BdToKSpp and $5050 < m(\KSpipi) < 5600\mevcc$ for \BdToKSpipi. For the sample optimised for the \BsToKSpp search, the fit is performed within the mass range $5000 < m(\KSpp) < 5600\mevcc$ on two subsamples, categorised by the \KS reconstruction type. While separate event selections are applied to the 2011–2012, 2015–2016, and 2017–2018 datasets, they share the same mass fits and efficiency calculations.

For both the \BdsToKSpp and the normalisation channel 
\BdToKSpipi decays, the fit model includes components for the signal, partially reconstructed background, and combinatorial background. The probability density function for the 
\Bd and \Bs signal peaks consists of two double-sided Crystal Ball (DSCB) functions~\cite{Skwarnicki:1986xj}. In the fit, the \Bd peak position and \Bds mass resolution are allowed to vary freely,  while the \Bs peak position is constrained as \mbox{$\mu(\Bs) = \mu(\Bd) + \Delta m$}, with \mbox{$\Delta m = 87.26\mevcc$} being the known mass difference between the \Bd and \Bs mesons~\cite{PDG2024}. The tail parameters of the DSCB functions for \Bz and \Bs are fixed to values obtained from their respective simulations.

The partially reconstructed background arises from \bquark-hadron decays into final states containing \KShh (where $h = \pi, \proton$) along with at least one unreconstructed particle. As a result, its invariant-mass distribution has a broad shape, peaking at values lower than the parent \bquark-hadron mass. Its shape is modelled using an ARGUS function convolved with a Gaussian function~\cite{ALBRECHT1990278}, where the curvature of the ARGUS function is fixed, while the slope, the threshold, and the Gaussian mean and width parameters are allowed to vary. The combinatorial background component, arising from combinations of unrelated particles, is modelled by an exponential function, with slope parameter allowed to vary.

Background contributions from \bquark-hadron decays that have a topology similar to the signal channel, but where final-state hadrons (\pip or \Kp) are misidentified, are also considered in the fits. The contribution from double misidentification is found to be negligible due to its small yield, and therefore only the singly misidentified components are included in the fit to the normalisation mode \BdToKSpipi. These components are modelled using a DSCB function, with the tail parameters fixed from simulation, while the peak position and width are allowed to vary. In the fit, the \BdsToKSKpi yield is constrained to the yield of the \BdToKSpipi decays with a fraction that incorporates the known branching fraction ratio, the efficiency ratio from simulation, and the misidentification rates measured in data control samples.

\begin{figure}[!hbtp]
  \begin{center}
    \includegraphics[scale=0.38]{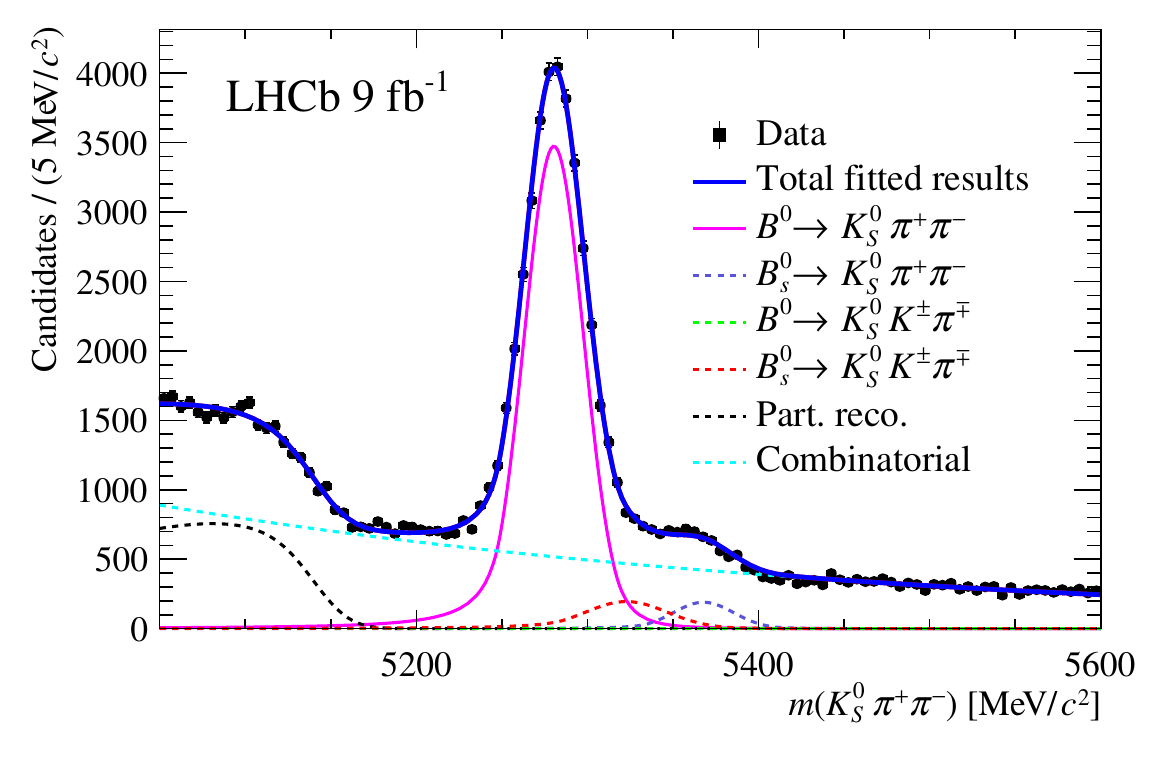}
    \includegraphics[scale=0.38]{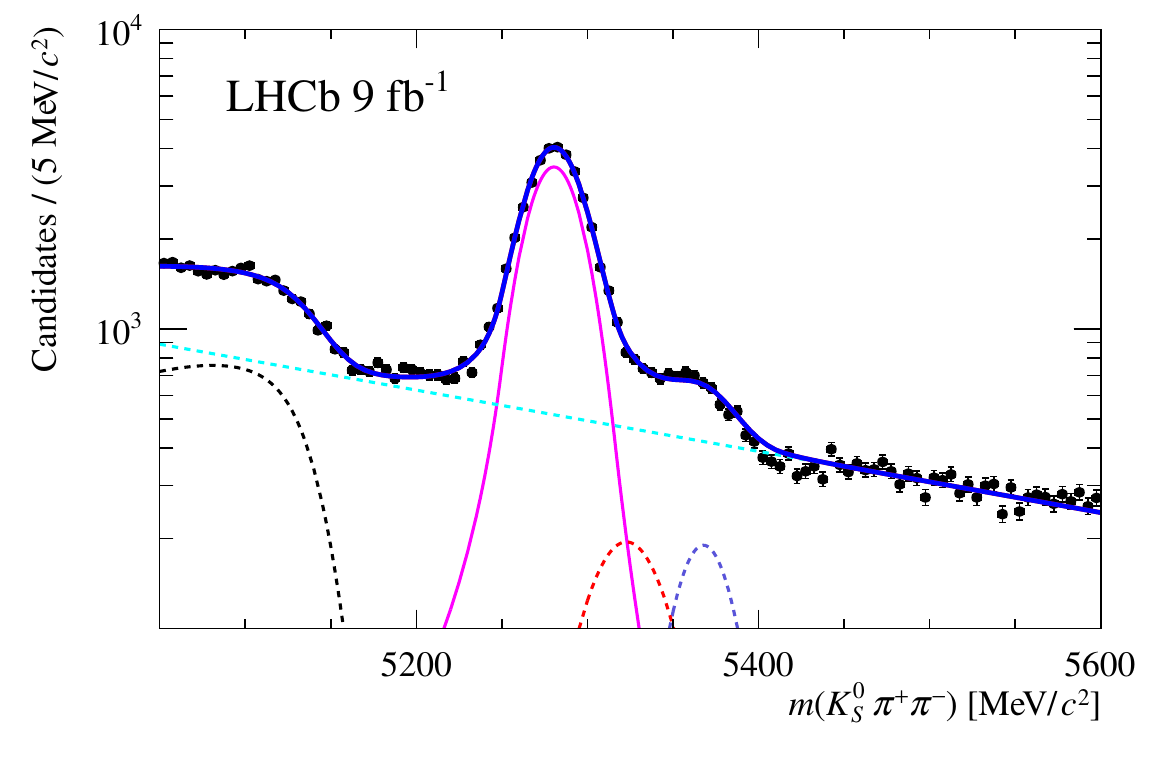}
    \includegraphics[scale=0.38]{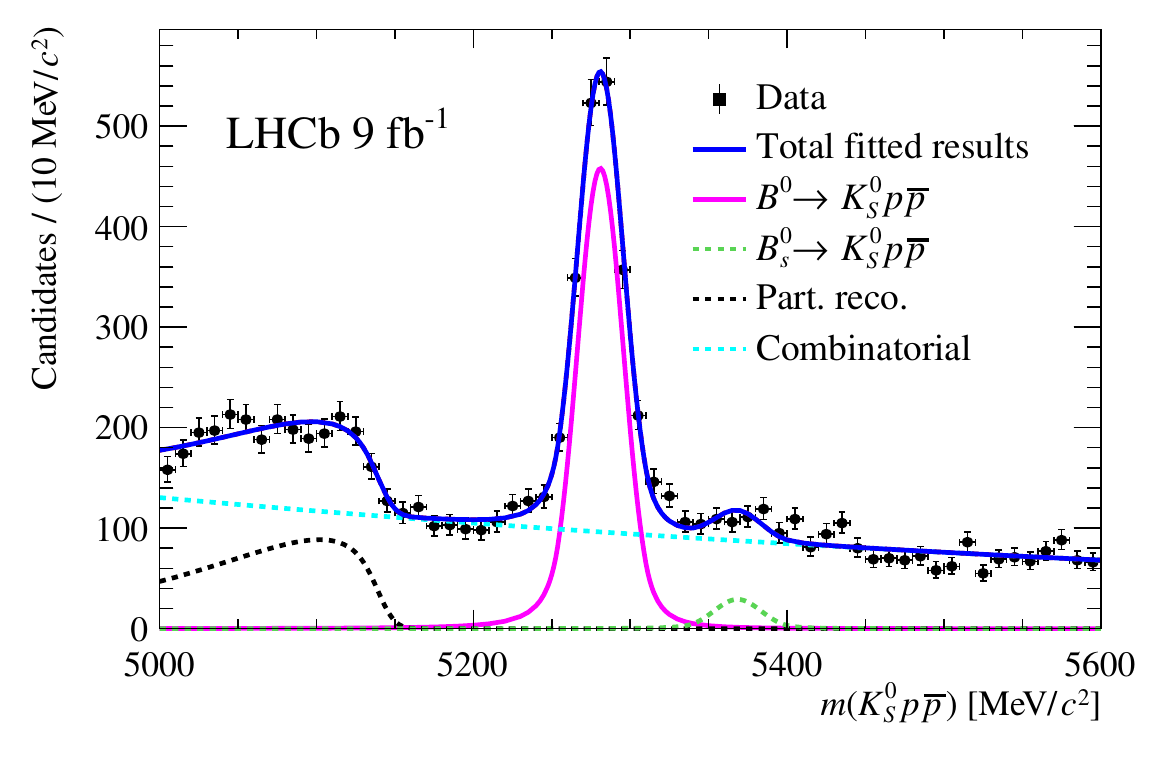}
    \includegraphics[scale=0.38]{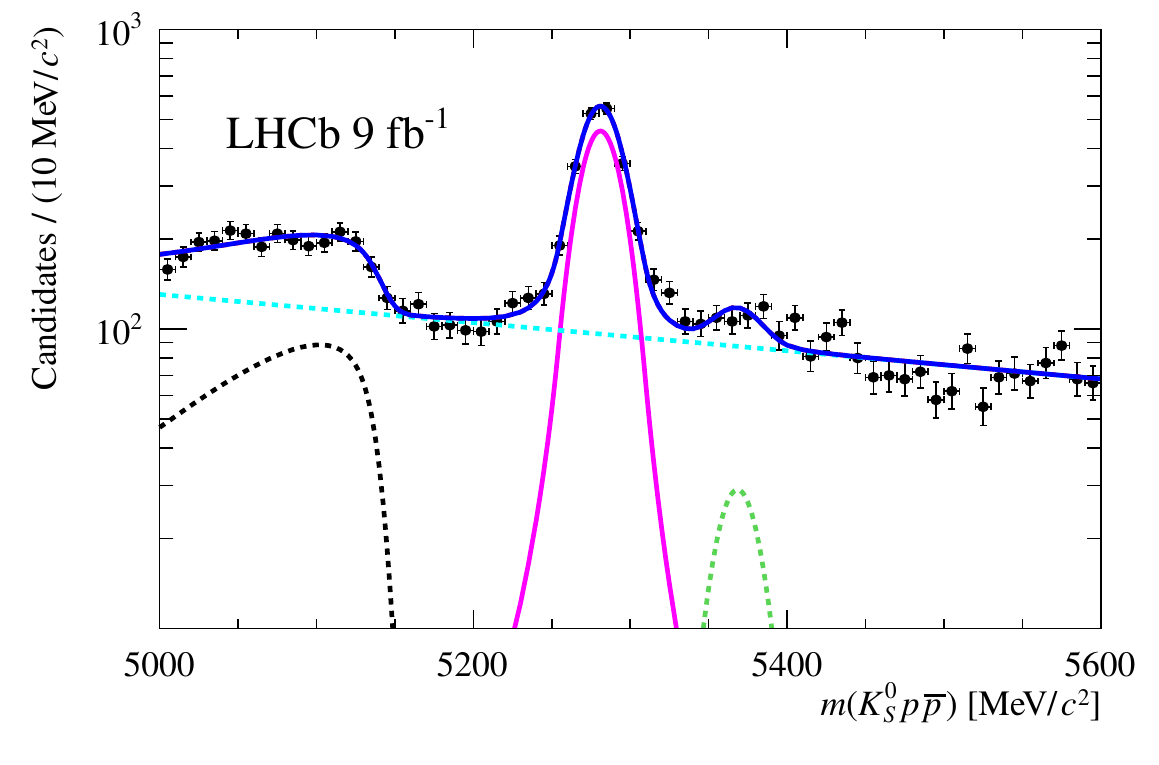}
    \includegraphics[scale=0.38]{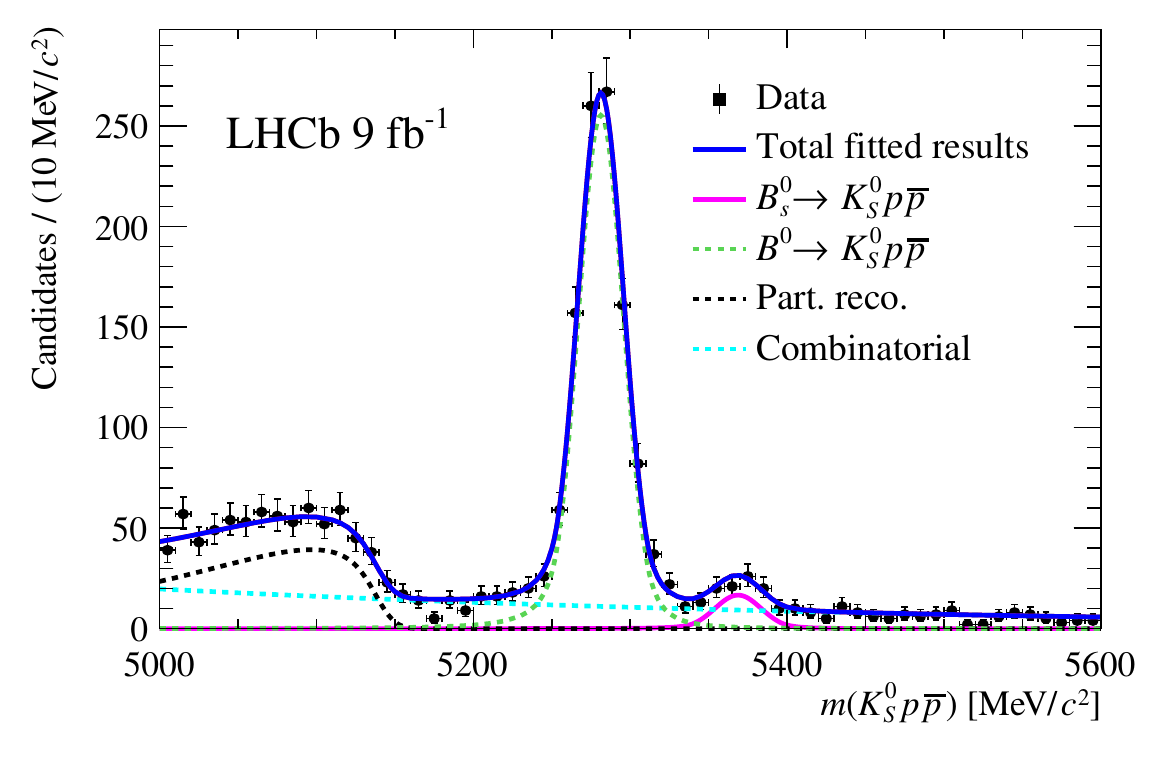}
    \includegraphics[scale=0.38]{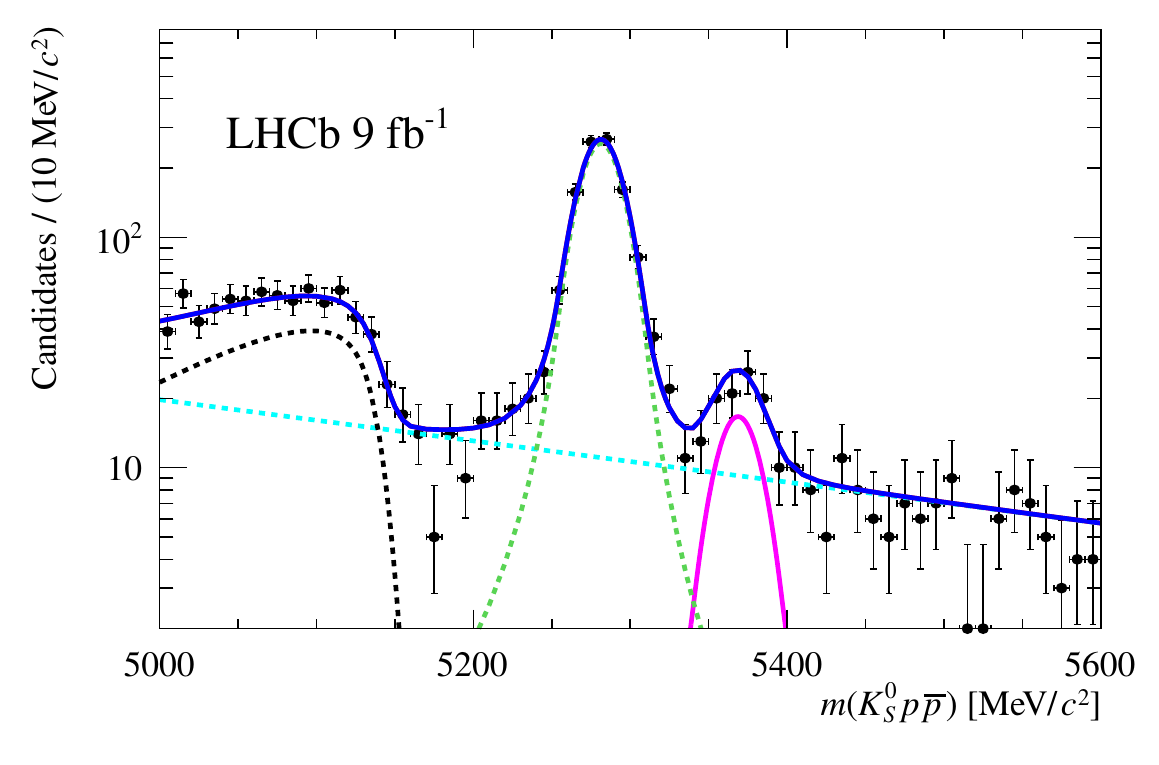}
  \end{center}
  \caption{Invariant-mass distributions of \KShh candidates that pass all selection criteria, combining all subsamples, are shown (left) on a linear scale and (right) on a logarithmic scale. Each row corresponds to a given decay mode: (top) \BdToKSpipi, (middle) \BdToKSpp, (bottom) \BsToKSpp. Fit results are also shown.}
\label{fig:fit_total_restlt_data}
\end{figure}

The individual fits of subsamples are presented in Appendix~\ref{sec:Fits:subsamples}. The sum of signal yields from the individual fits are $1\,791 \pm 52$ for \BdToKSpp and $32\,145 \pm 230$ for \BdToKSpipi. A peaking structure is observed in the $m(\KS \proton\antiproton)$ distribution near the known \Bs meson mass. The sum of the signal yields, determined using the selection criteria optimised for the \BsToKSpp search, is $66 \pm 12$. Based on statistical uncertainty alone, the signal significance is $7.1\sigma$, as calculated using Wilks' theorem~\cite{Wilks:1938dza}. When systematic uncertainties are also taken into account, the significance decreases to $5.6\sigma$, as discussed in Sec.~\ref{sec:Results:conclusion}. For visualisation, the fit results for the full dataset are shown in Fig.~\ref{fig:fit_total_restlt_data}.
%%%%%%%%%%%%%%%%%%%%%%%%%%%%%%%%%%%%%%%%%%%%%%%%%%%%%%%%%%%%%%%%%%
%%%%%%%%%%%%%%%%%%%%%%%%%%%%%%%%%%%%%%%%%%%%%%%%%%%%%%%%%%%%%%%%%%

%%%%%%%%%%%%%%%%%%%%%%%%%%%%%%%%%%%%%%%%%%%%%%%%%%%%%%%%%%%%%%%%%%%%%%%%%%%%%%%%%%%%%%%%%%%%%%%%%%
\section{Branching fraction measurement}
\label{sec:Determination:of:the:efficiencies}
The branching fraction  of \BdsToKSpp decays is determined relative to the normalisation channel \BdToKSpipi through the ratio
%%%%%%%%%%%%%%%%%%%%%%%%%%%%%%%%%%%%%%%%%%%%%%%%%%%%%%%%%%%%%%%%%%
\begin{equation}
\label{eq:BR}
\frac{\BR(\BdsToKSpp)}{\BR(\BdToKSpipi)} = \frac{\varepsilon^{\rm sel}(\BdToKSpipi)}{\varepsilon^{\rm sel}(\BdsToKSpp)} \times \frac{N(\BdsToKSpp)}{N(\BdToKSpipi)} \times \frac{f_d}{f_{d(s)}},
\end{equation}
%%%%%%%%%%%%%%%%%%%%%%%%%%%%%%%%%%%%%%%%%%%%%%%%%%%%%%%%%%%%%%%%%%
where $\varepsilon^{\rm sel}(X)$ and $N(X)$ are the efficiencies and yields of the different channels, $X$, and $f_d$ and $f_s$ are the hadronisation fractions of a \bquark quark into a \Bd or \Bs meson, respectively. 
The efficiency term includes the effects of geometrical acceptance and trigger, reconstruction and selection; it is determined using simulation and dedicated calibration procedures. The ratio $f_s/f_d$ has been measured by the \lhcb experiment, using hadronic and semileptonic decays at different centre-of-mass energies~\cite{LHCb-PAPER-2020-046}. Taking into account  the integrated luminosity and the production cross-section at each centre-of-mass energy, a weighted average value of $0.2486 \pm 0.0078$ is obtained.

Three-body hadronic decays, such as \BdsToKShh, exhibit a complex phase-space structure that can be visualised in a Dalitz plot~\cite{Dalseno_2009}. The decay amplitude can be influenced by intermediate quasi-two-body resonances and broad, nonresonant contributions that overlap and interfere. This nonuniform distribution of events typically results in higher probability densities near the kinematic boundaries of the Dalitz plot.
The overall efficiency of the reconstruction and selection depends on the Dalitz structure of the decay. To simplify the description of the phase space, and to avoid edge effects, the Dalitz plot is transformed into a square representation~\cite{LHCb-PAPER-2018-045}. This transformation ensures a more uniform distribution of candidates near the kinematic boundaries, where resonances are concentrated and efficiency variations are significant. Along with this transformation, the simulation is weighted to match the data. Background-subtracted Dalitz plots are obtained using the \sPlot method~\cite{Pivk:2004ty}, where each candidate is assigned a weight derived from fits to the $m(\KSpp)$ and $m(\KSpipi)$ distributions in Sec.~\ref{sec:Fit:model}. The selection efficiency is then corrected using signal weights from data and the per-event efficiency from weighted simulation, giving 
%%%%%%%%%%%%%%%%%%%%%%%%%%%%%%%%%%%%%%%%%%%%%%%%%%%%%%%%%%%%%%%%%%
\begin{equation}
\label{eq:sPlots}
\varepsilon^{\rm sel} =\frac{\sum_{e}W_e}{\sum_{e}\frac{W_e}{\varepsilon_{e(j)}}},
\end{equation}
%%%%%%%%%%%%%%%%%%%%%%%%%%%%%%%%%%%%%%%%%%%%%%%%%%%%%%%%%%%%%%%%%%
where $W_e$ is the signal weight associated with the candidate $e$ and $\varepsilon_{e(j)}$ its efficiency, taken from the corresponding bin $j$ in the square Dalitz plot, which is split into a uniform $5 \times 5$ grid. The value of $\varepsilon_{e(j)}$ varies in the range 0.001--0.005 for \BdToKSpp, 0.001--0.003 for \BdToKSpipi and 0.0003--0.0013 for \BsToKSpp decays, with a smooth and well-controlled overall behaviour.

The PID efficiencies and misidentification rates are derived from simulations based on the kinematic properties of the tracks in the decay process. These are then corrected, exploiting the PID response in calibration data. A weighting procedure is performed in bins of momentum, pseudorapidity, and event multiplicity, accounting for kinematic correlations between the tracks~\cite{LHCb-PUB-2016-021}. 
Differences in tracking efficiency are corrected as a function of the track momentum and pseudorapidity using data control channels.
Differences in hardware trigger response between data and simulation are determined using calibration samples of \DstarToDpiToKpi and \LzToppi decays where the trigger requirement is satisfied independently of the decay products. The trigger response is analysed based on the hadron charge, \pt, calorimeter region, and cluster overlap of final-state particles from signal decays, with a correction for the average underlying energy, which includes contributions from multiple interactions and diffuse energy deposits~\cite{LHCb-DP-2012-004,CalvoGomez:2042173,LHCb-DP-2020-001}.

The efficiency correction due to mass vetoes is estimated using simulation, which implements a uniform phase-space decay model. Therefore, the measured results can be directly compared with those from other experiments, which are determined across the entire  \proton\antiproton mass range. 
Finally, the efficiency ratio between \BdToKSpipi and \BdToKSpp decays is $1.02 \pm 0.02$, while that between \BdToKSpipi and \BsToKSpp decays is $2.23 \pm 0.12$. The significantly larger ratio observed in the latter case is primarily attributed to differences in the selection criteria, with the most notable contribution arising from the distinct BDT requirements applied to the three decay modes. The uncertainties stem from the limited size of the signal simulation. These efficiency ratios are calculated by combining the values determined on each data sample.

%%%%%%%%%%%%%%%%%%%%%%%%%%%%%%%%%%%%%%%%%%%%%%%%%%%%%%%%%%%%%%%%%%%%%%%%%%%%%%%%%%%%%%%%%%%%%%%%%%
\section{Systematic uncertainties}
\label{sec:Systematic:uncertainties}
The \BdsToKSpp branching fractions are measured relative to the decay \BdToKSpipi, as expressed in Eq.~\ref{eq:BR}. Since similar selections and fit procedures are used for decays in the numerator and denominator, several potential systematic effects are expected to cancel in the ratio. The remaining uncertainties are discussed in this section and categorised based on their influence on either the yield measurements ($N$) or the efficiency ($\varepsilon^{\rm sel}$). The uncertainty on ${f_s}/{f_d}$ is also included where appropriate. The total, statistical and systematic uncertainties are summarised in Table~\ref{tab:sys:total}.
%%%%%%%%%%%%%%%%%%%%%%%%%%%%%%%%%%%%%%%%%%%%%%%%%%%%%%%%%%%%%%%%%%

The systematic uncertainty associated with the choice of fit model is evaluated by comparing signal yields obtained from the baseline and alternative fit models, with variations applied to the signal, partially reconstructed, misidentified, and combinatorial background components. 
The difference in signal yields between the two models is fitted with a Gaussian function, and the systematic uncertainty is taken as the quadrature sum of the Gaussian mean and width.

For each alternative model, one fit component is varied at a time. For the signal and  misidentified background models, the baseline DSCB functions are replaced with a sum of a Student's t-distribution and a Gaussian function. The exponential function modelling the combinatorial background is replaced by a second-order Chebyshev polynomial function. To estimate the impact of partially reconstructed background, the fit is performed in the restricted invariant-mass range of 5150--5600\mevcc, where its contribution is reduced. 

The accuracy of the efficiency determination is limited by the size of the signal simulation samples, which is propagated as a systematic uncertainty.
Additionally, the effect of the choice of binning scheme used to subdivide the square Dalitz plot is estimated based on the spread of the average efficiencies determined from various alternative binning schemes.

The procedure to evaluate the efficiencies of the PID requirements, described in Sec.~\ref{sec:Fit:model}, 
uses calibration tracks that systematically differ from the signal tracks. While the binning procedure attempts to mitigate these differences, there could be some remaining systematic effect. This is addressed by considering different ensembles of kinematical binning schemes to determine the efficiency. An overall 1\% systematic uncertainty is assigned to quantify any bias due to the procedure and is added in quadrature with the statistical uncertainties originating from the finite size of the simulation samples. 
A relative systematic uncertainty on the tracking corrections is assigned for the data-taking periods of Run~1 and Run~2, accounting for differences in hadronic interactions of protons and pions with the detector material~\cite{LHCb-PAPER-2019-034}.

Possible sources of systematic uncertainty related to the efficiency estimation of the hardware trigger have been studied. Two sources are identified: one arising from the imperfect simulation of the rate at which overlapping tracks in the hadron calorimeter merge into a single hadron trigger candidate, and the other from the choice of the data calibration sample. The first uncertainty is evaluated as the difference in the trigger efficiency correction with and without the overlapping cluster corrections. For the second, correction factors are determined using a sample of reconstructed \BdTojpsiKpi decays, and the difference between the factors obtained from the two calibration samples is assigned as the systematic uncertainty. These two uncertainties are treated as independent and are summed in quadrature to obtain the total systematic uncertainty associated with the hardware trigger efficiency. The differences observed between the \BdToKSpp and \BsToKSpp channels are attributed to the limited sample size and to possible residual double counting effects, which are expected to be small.

The uncertainty associated with the BDT selection is evaluated by varying the requirement and recalculating the efficiency and yields. The resulting deviation in the branching fractions is assigned as an uncertainty. Similarly, the uncertainty from the choice of mass veto regions, used to exclude charm and charmonium contributions, is assessed by varying their size and evaluating the selection performance again. The resulting changes are used to determine the deviation in the branching fractions, which is assigned as the corresponding uncertainty. It is important to note that the efficiency is evaluated over the full range of the \proton\antiproton and \KS\proton invariant masses. While the vetoed regions do contain a small amount of signal, they represent only a tiny portion of the phase space. As a result, excluding these regions has a minimal effect on the overall efficiency and the final results.

The uncertainty in the \Bd and \Bs lifetimes affects the selection efficiency. For \BdToKSpp and \BdToKSpipi decays, the effect on the ratio of branching fractions is negligible. The width difference between the \BsH and \BsL states leads to a lifetime variation of up to 5\%~\cite{PDG2024,De_Bruyn_2012}. Following Ref.\cite{Aaltonen_2011}, the systematic uncertainty is evaluated by weighting the default \BsToKSpp decay time distribution and comparing it to variations of $\pm5\%$. The resulting maximum deviation in the \Bs efficiency is quoted in Table\,\ref{tab:sys:total}. 

The systematic uncertainties are combined while accounting for their correlations. The correlation coefficients are 0\% for the sample size and mass veto systematics, 50\% for the fit model, PID, binning, tracking, and L0 trigger systematics, and 100\% for the BDT selection. The `L0 Trigger' uncertainty is the dominant contribution.
%%%%%%%%%%%%%%%%%%%%%%%%%%%%%%%%%%%%%%%%%%%%%%%%%%%%%%%%%%%%%%%%%%
\begin{table}[tb]
\caption{Summary of the relative statistical and systematic uncertainties on the branching fraction ratios, expressed in percent. The total systematic uncertainty represents the quadrature sum of all individual systematic contributions.}
\centering
\begin{tabular}{c|c|c}
\hline \hline                       & $\frac{\BR(\Bd  \to \KS\proton\antiproton)}{\BR(\Bd  \to \KS \pip\pim)}$ & $ \frac{\BR(\Bs  \to \KS \proton\antiproton)}{\BR(\Bd  \to \KS\pip\pim)}$ \\ \hline     
   Statistical    &   3.0\%  & 18.4\%             \\     \hline
   Fit model     &   1.0\%  & 4.7\%                         \\ 
   Simulation sample size  &   1.7\%  & 5.5\%             \\
   Binning       &   2.4\%  & 2.6\%                         \\
    PID           &   1.1\%  & 1.2\%                         \\
   Tracking      &   0.5\%  & 0.5\%                         \\
   L0 trigger    &   3.6\%  & 5.5\%                         \\
   BDT selection &   0.9\%  & 1.2\%                         \\
   Mass veto     &   0.7\%  & 1.8\%                         \\
   \Bs lifetime & --- & 3.1\%              \\ \hline
   Total systematic        &   5.0\%  & 9.8\%            \\ \hline
   ${f_s}/{f_d}$ &  ---   & 2.2\%                  \\ \hline
    \hline
    \end{tabular}
    \label{tab:sys:total}
    \end{table}
%%%%%%%%%%%%%%%%%%%%%%%%%%%%%%%%%%%%%%%%%%%%%%%%%%%%%%%%%%%%%%%%%%
\section{Results and conclusion}
\label{sec:Results:conclusion}
A search for the \BsToKSpp decay and measurements of the branching fraction ratios of \BdsToKSpp decays are performed, using a $pp$ collision data sample recorded with the \lhcb detector from 2011 to 2018, corresponding to an integrated luminosity of 9\invfb. The decay \BsToKSpp is observed for the first time, with a significance of 5.6 standard deviations, including both statistical and systematic uncertainties.

The branching fractions of the two decay modes are measured relative to the \BdToKSpipi decay. The branching fraction ratios are determined independently for two \KS reconstruction categories and different data-taking periods for \BdToKSpp, and found to be consistent between all subsamples.
%With consideration of systematic uncertainty correlations among different subsamples,
Combining these results using the Best Linear Unbiased Estimate method~\cite{Nisius_2014}, which takes into account the correlations of systematic uncertainties among different subsamples, yields:
%%%%%%%%%%%%%%%%%%%%%%%%%%%%%%%%%%%%%%%%%%%%%%%%%%%%%%%%%%%%%%%%%%
\begin{align}
    \frac{\BR(\BdToKSpp)}{\BR(\BdToKSpipi)} &= \BdKSppResult, \nonumber \\
    \frac{\BR(\BsToKSpp)}{\BR(\BdToKSpipi)} &= \BsKSppResult, \nonumber \\
    \frac{\BR(\BsToKSpp)}{\BR(\BdToKSpp)} &= \BsKSppResultBd, \nonumber 
\end{align}
where the first uncertainty is statistical, the second systematic in each case, and the third originates from the ratio of fragmentation fractions of the \Bs and \Bd mesons, ${f_s}/{f_d}$. Using the known branching fraction of the normalisation channel, 
\(\BR(\BdToKzpipi) = \mbox{$\BdKzpipiResultBF$} \)~\cite{PDG2024}, 
the absolute branching fractions of \BdToKzpp and \BsToKzpp decays are determined to be 
\begin{equation}\nonumber
    \begin{aligned}
        \label{fom:BR}
        \BR(\BdToKzpp) &= \BdKzppResultBF, \\
        \BR(\BsToKzpp) &= \BsKzppResultBF,
    \end{aligned}
\end{equation}
where the third and fourth uncertainties are due to the uncertainties of the branching fraction of the normalisation channel, and ${f_s}/{f_d}$, respectively. The measured \BdToKzpp branching fraction is consistent with and more precise than the world average value \mbox{$\BdKzppResultBFPDG$}~\cite{PDG2024}.

%%%%%%%%%%%%%%%%%%%%%%%%%%%%%%%%%%%%%%%%%%%%%%%%%%%%%%

% Do not include this in any draft (just for information in the template)
% \input{acknowledgements_intro}
% Comment this in for paper drafts; do not include this in analysis note, conference and figure reports
\section*{Acknowledgements}
%
% These Acknowledgements valid from 3-May-2019
%
\noindent We express our gratitude to our colleagues in the CERN
accelerator departments for the excellent performance of the LHC. We
thank the technical and administrative staff at the LHCb
institutes.
We acknowledge support from CERN and from the national agencies:
ARC (Australia);
CAPES, CNPq, FAPERJ and FINEP (Brazil); 
MOST and NSFC (China); 
CNRS/IN2P3 (France); 
BMBF, DFG and MPG (Germany); 
INFN (Italy); 
NWO (Netherlands); 
MNiSW and NCN (Poland); 
MCID/IFA (Romania); 
%MSHE (Russia); 
MICIU and AEI (Spain);
SNSF and SER (Switzerland); 
NASU (Ukraine); 
STFC (United Kingdom); 
DOE NP and NSF (USA).
%%%%%%%%%%%%%%%%%%%%%%%%%%%%%%%%%%%%%%%%%%%%%
We acknowledge the computing resources that are provided by ARDC (Australia), 
CBPF (Brazil),
CERN, 
IHEP and LZU (China),
IN2P3 (France), 
KIT and DESY (Germany), 
INFN (Italy), 
SURF (Netherlands),
Polish WLCG (Poland),
IFIN-HH (Romania), 
%RRCKI and Yandex LLC (Russia), 
PIC (Spain), CSCS (Switzerland), 
and GridPP (United Kingdom).
%%%%%%%%%%%%%%%%%%%%%%%%%%%%%%%%%%%%%%%%%%
We are indebted to the communities behind the multiple open-source
software packages on which we depend.
%%%%%%%%%%%%%%%%%%%%%%%%%%%%%%%%%%%%%%%%%%
Individual groups or members have received support from
%ARC and ARDC (Australia); % moved to national 16/01
Key Research Program of Frontier Sciences of CAS, CAS PIFI, CAS CCEPP, 
Fundamental Research Funds for the Central Universities,  and Sci.\ \& Tech.\ Program of Guangzhou (China);
Minciencias (Colombia);
EPLANET, Marie Sk\l{}odowska-Curie Actions, ERC and NextGenerationEU (European Union);
A*MIDEX, ANR, IPhU and Labex P2IO, and R\'{e}gion Auvergne-Rh\^{o}ne-Alpes (France);
%RFBR, RSF and Yandex LLC (Russia);
Alexander-von-Humboldt Foundation (Germany);
ICSC (Italy); 
%GVA, XuntaGal, GENCAT, Inditex, InTalent and Prog.~Atracci\'on Talento, CM (Spain);
Severo Ochoa and Mar\'ia de Maeztu Units of Excellence, GVA, XuntaGal, GENCAT, InTalent-Inditex and Prog.~Atracci\'on Talento CM (Spain);
SRC (Sweden);
the Leverhulme Trust, the Royal Society and UKRI (United Kingdom).

\newpage
{\noindent\normalfont\bfseries\Large Appendices}
\appendix
\section{Fit results of all subsamples}
\label{sec:Fits:subsamples}
\begin{figure}[!htb]
  \begin{center}
    \includegraphics[width=0.45\textwidth]{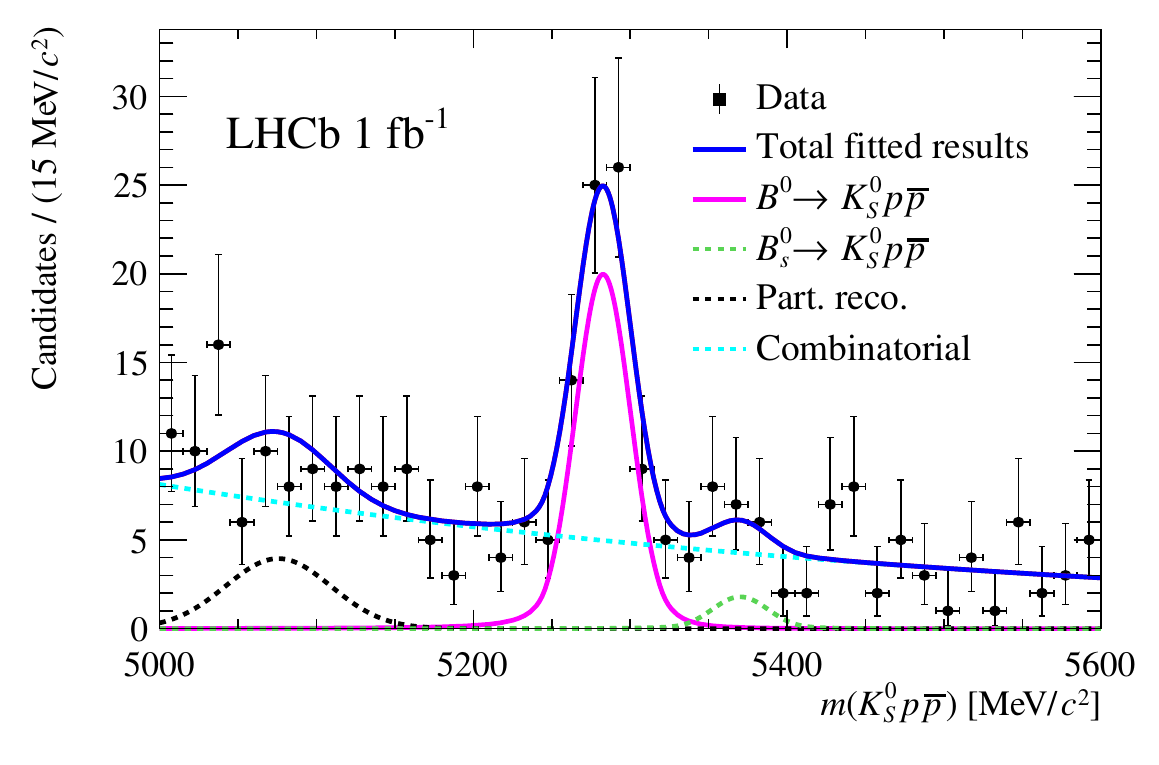}
    \includegraphics[width=0.45\textwidth]{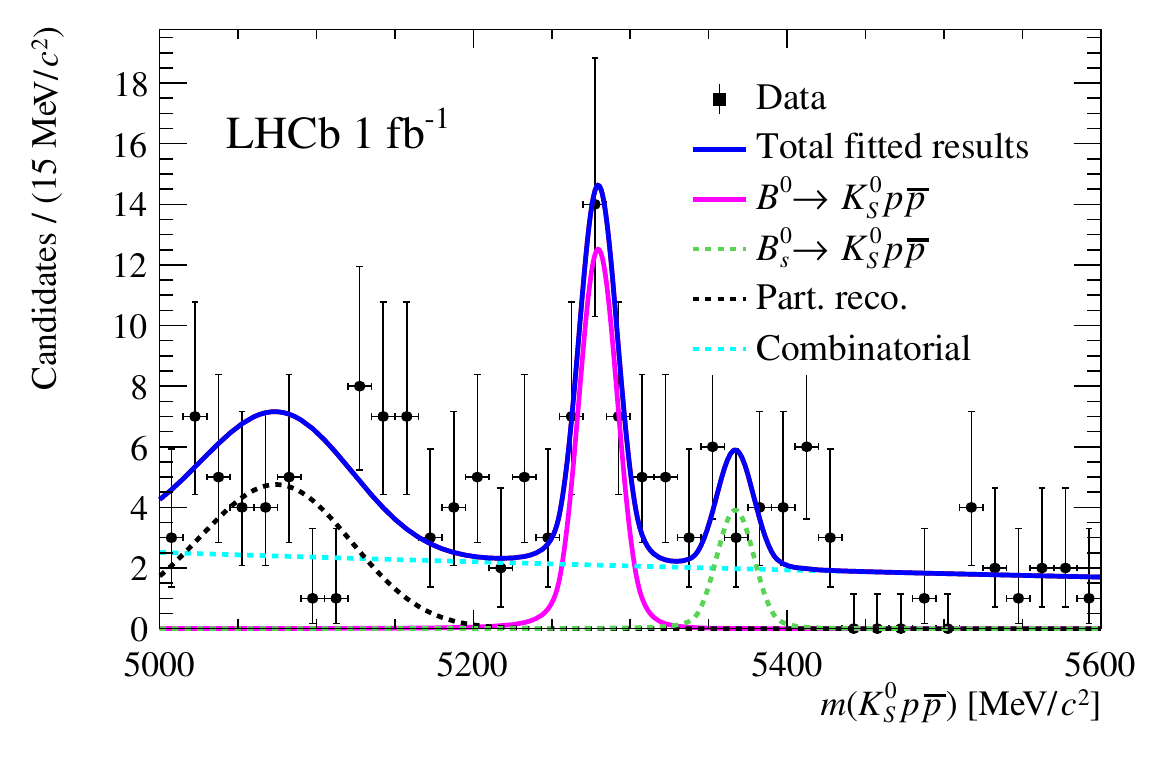}
    \includegraphics[width=0.45\textwidth]{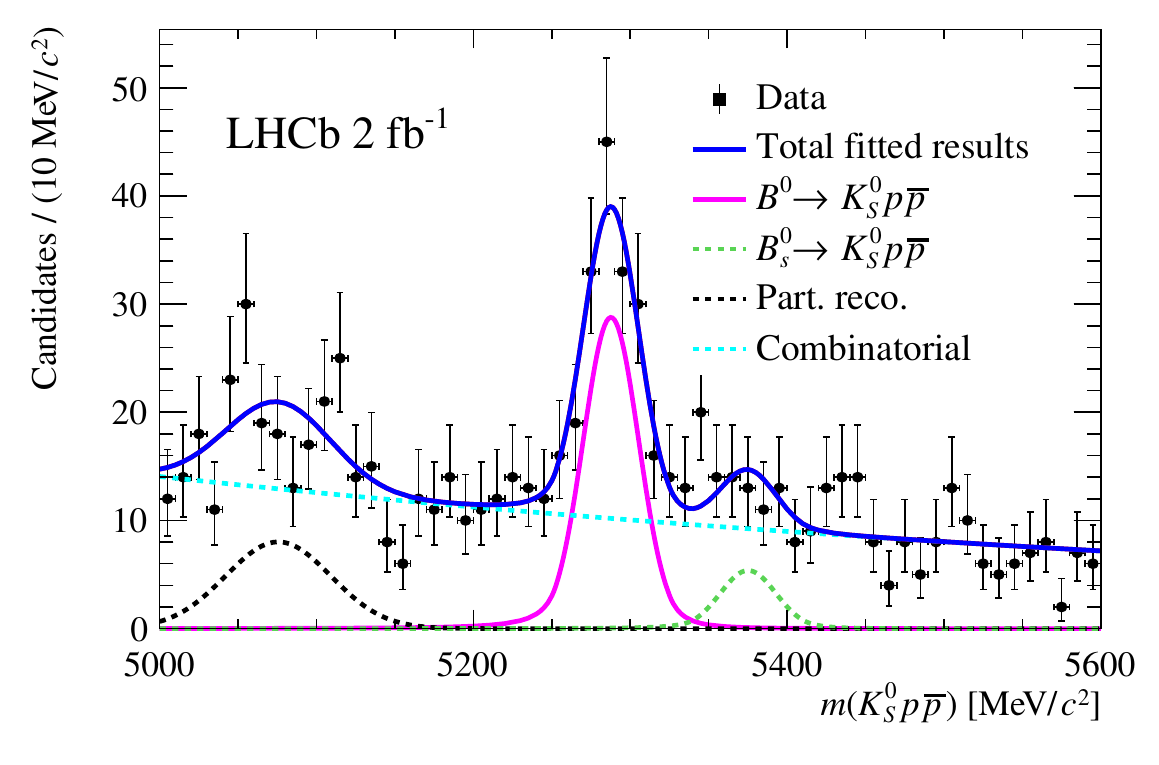}
    \includegraphics[width=0.45\textwidth]{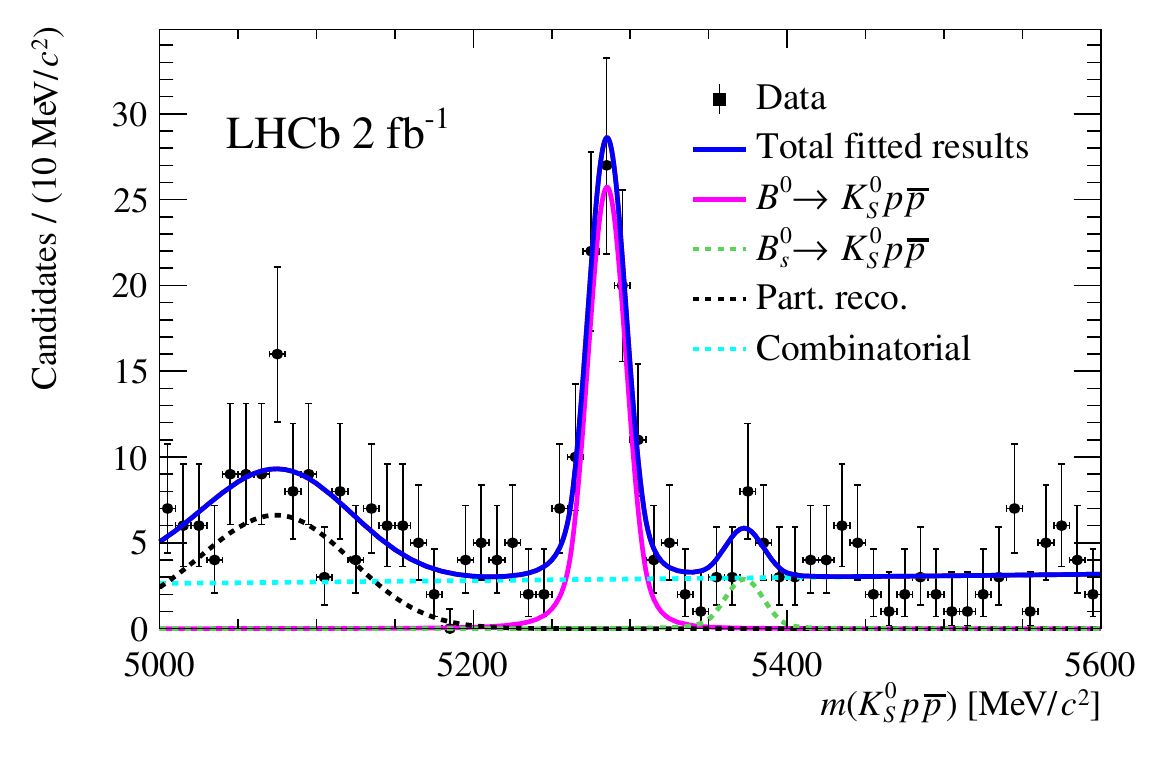}
  \end{center}
  \caption{Invariant-mass distributions of \KSpp candidates passing the \BdToKSpp selection. Data corresponding to (left) downstream and (right) long \KS reconstruction categories are shown for (top) the 2011 dataset and (bottom) the 2012 dataset. Fit results are also shown.}
  \label{Run1_BdKSpp}
\end{figure}

\begin{figure}[htb]
  \begin{center}
    \includegraphics[width=0.45\textwidth]{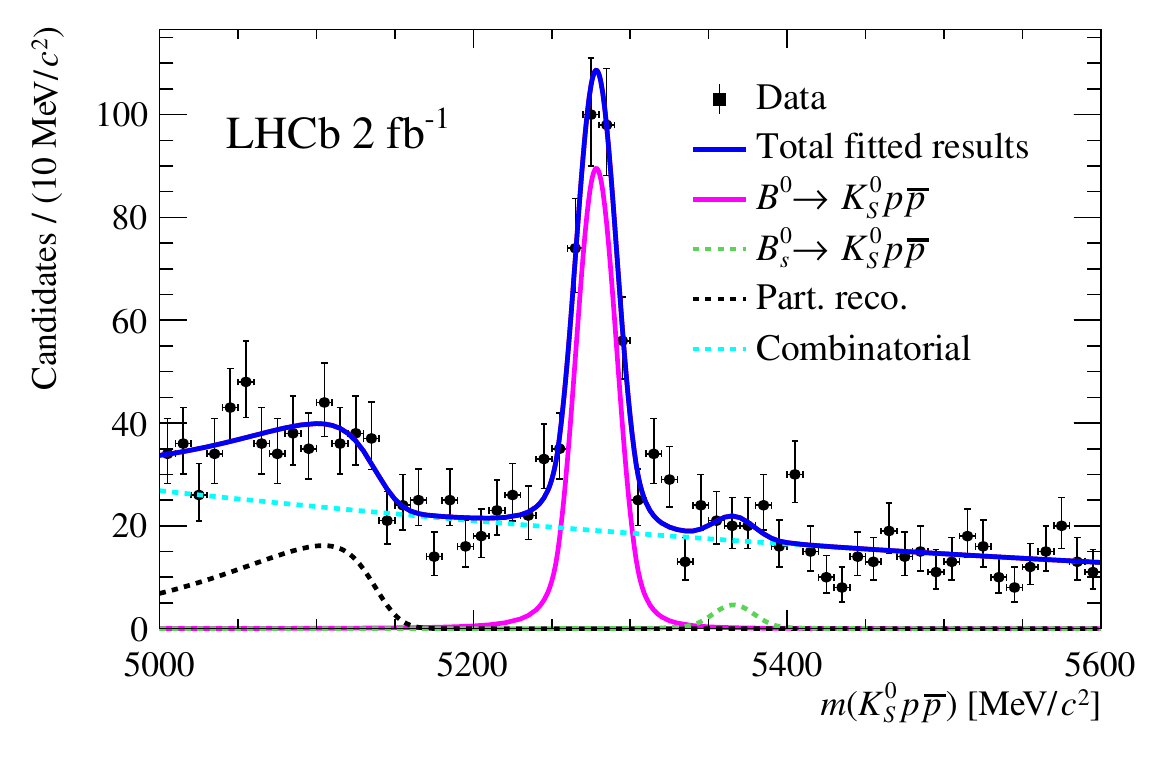}
    \includegraphics[width=0.45\textwidth]{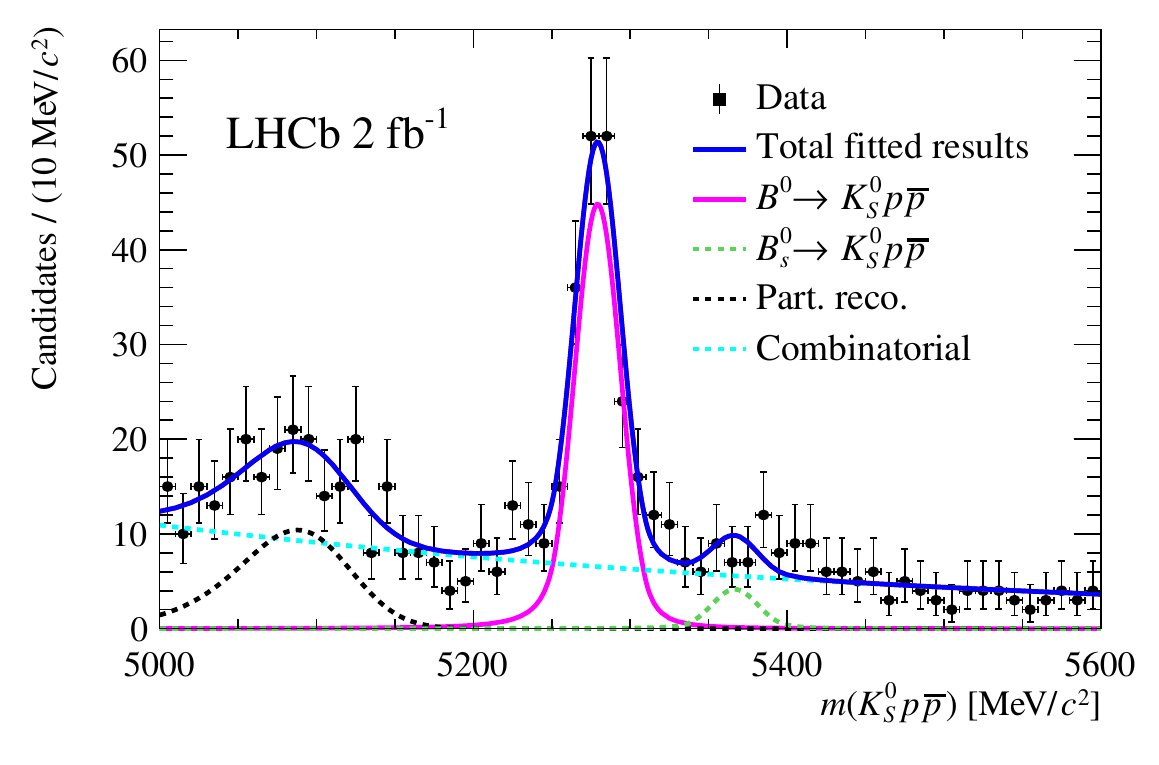}
    \includegraphics[width=0.45\textwidth]{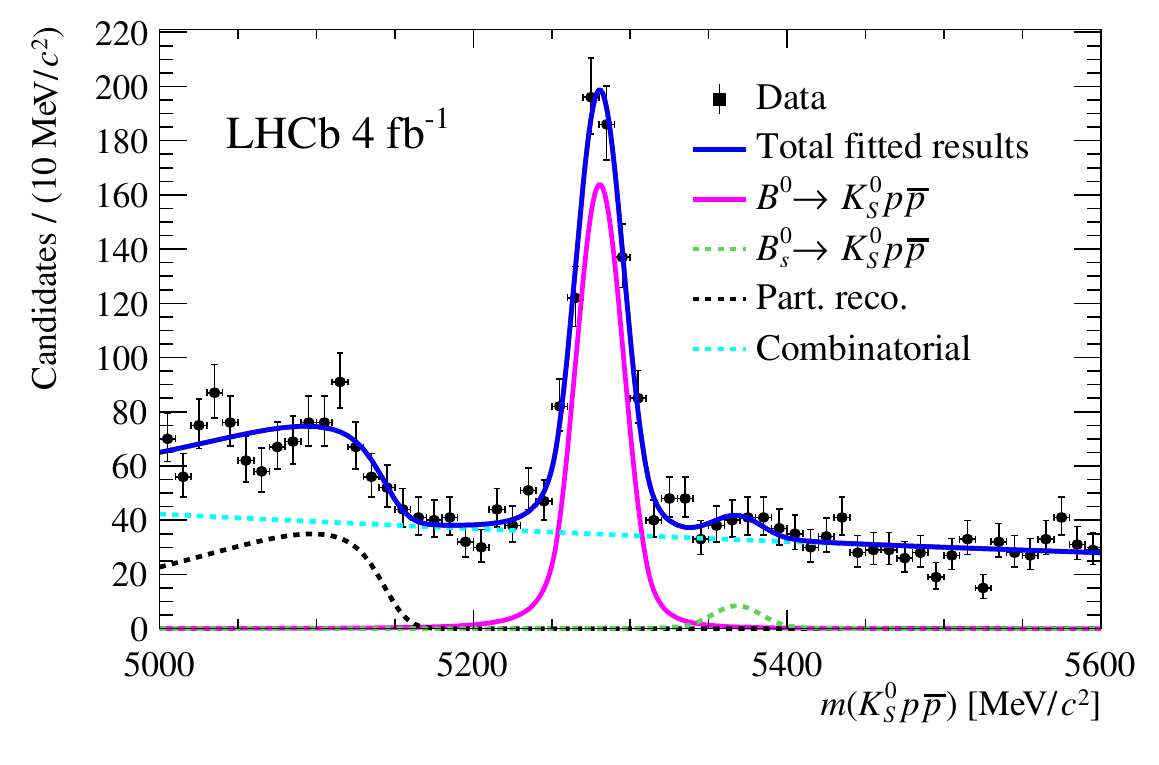}
    \includegraphics[width=0.45\textwidth]{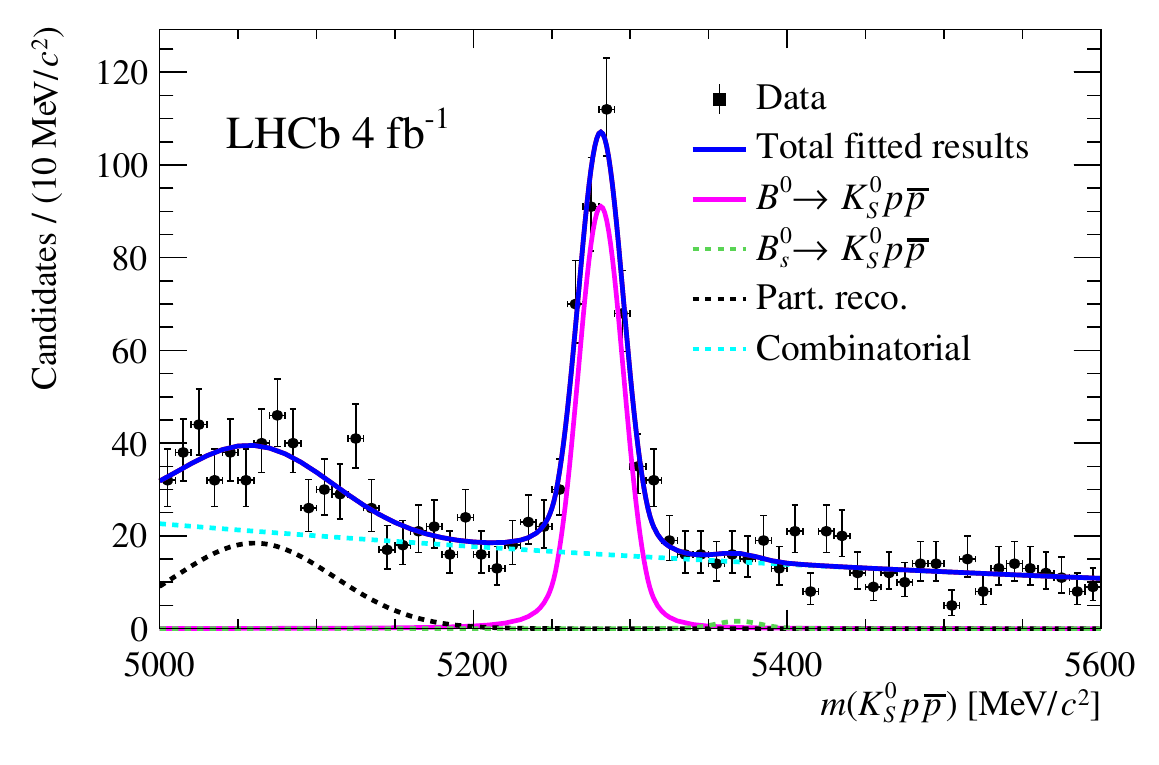}
  \end{center}
  \caption{Invariant-mass distributions of \KSpp candidates passing the \BdToKSpp selection. Data corresponding to (left) downstream and (right) long \KS reconstruction categories are shown for (top) the 2015--2016 dataset and (bottom) the 2017--2018 dataset. Fit results are also shown.}
  \label{Run2_BdKSpp}
\end{figure}

\begin{figure}[htb]
  \begin{center}
    \includegraphics[width=0.45\textwidth]{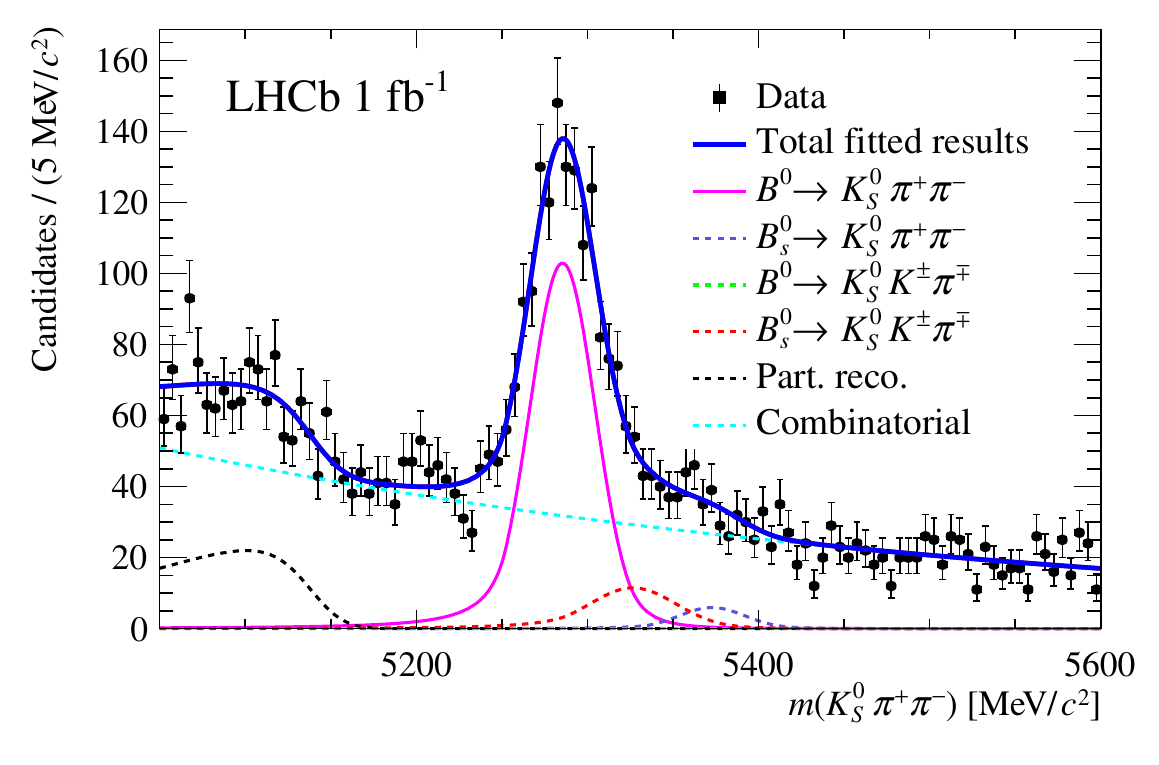}
    \includegraphics[width=0.45\textwidth]{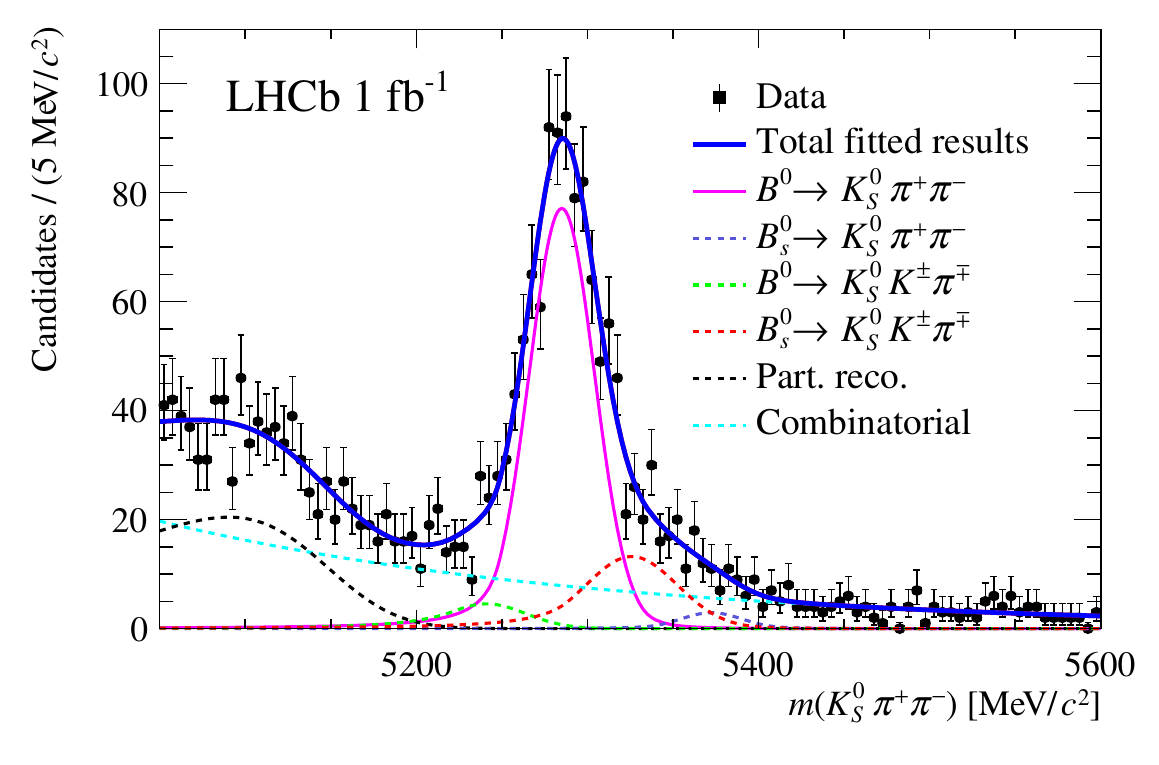}
    \includegraphics[width=0.45\textwidth]{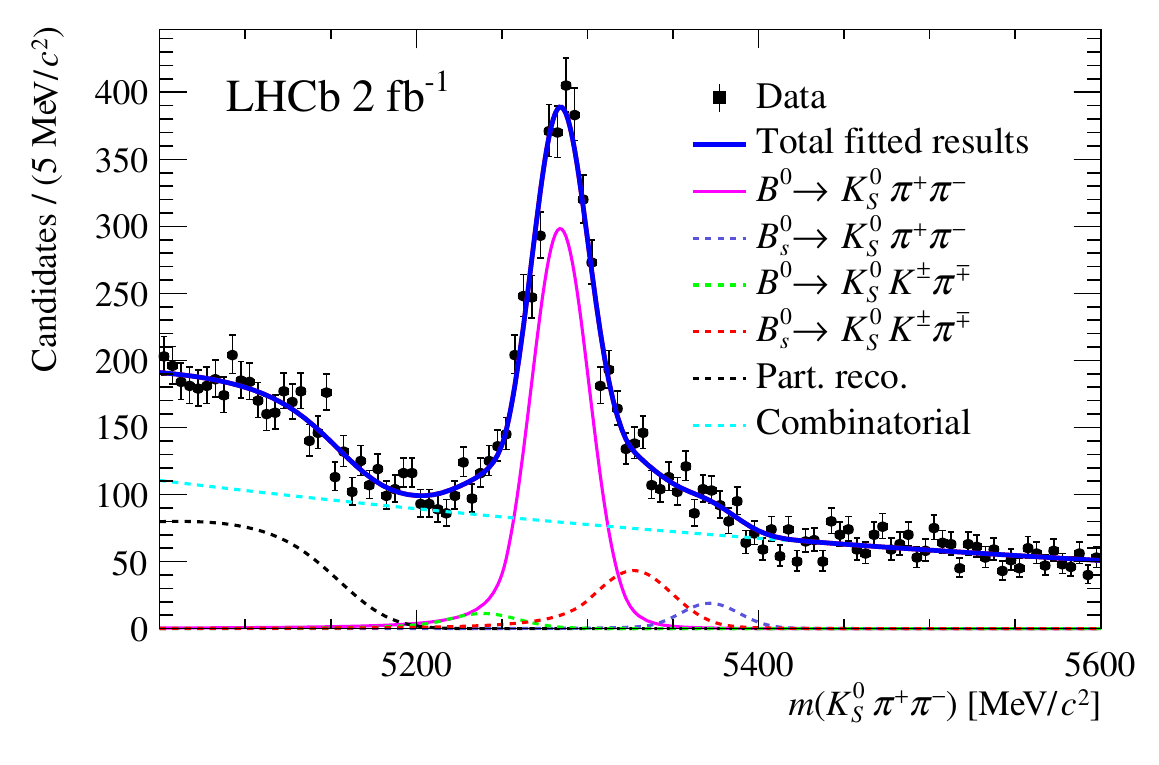}
    \includegraphics[width=0.45\textwidth]{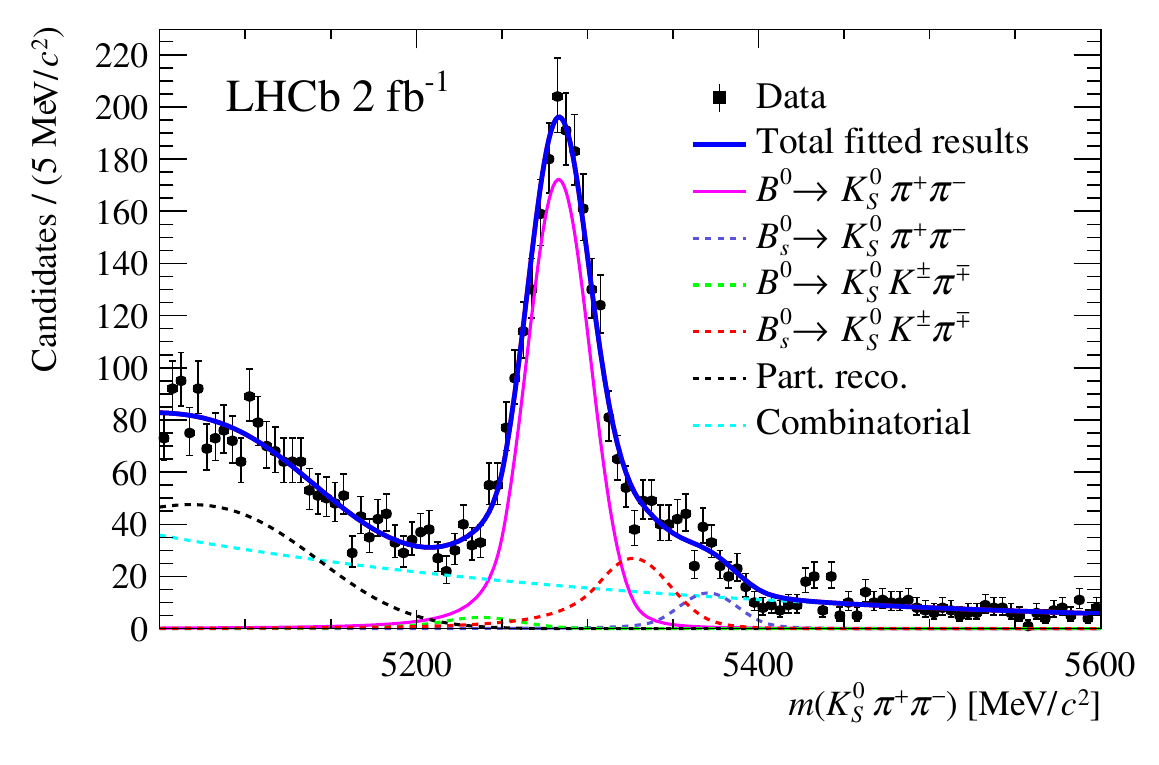}
  \end{center}
  \caption{Invariant-mass distributions of \KSpipi candidates passing the \BdToKSpipi selection. Data corresponding to (left) downstream and (right) long \KS reconstruction categories are shown for (top) the 2011 dataset and (bottom) the 2012 dataset. Fit results are also shown.}
  \label{Run1_BdKSpipi}
\end{figure}

\begin{figure}[htb]
  \begin{center}
    \includegraphics[width=0.45\textwidth]{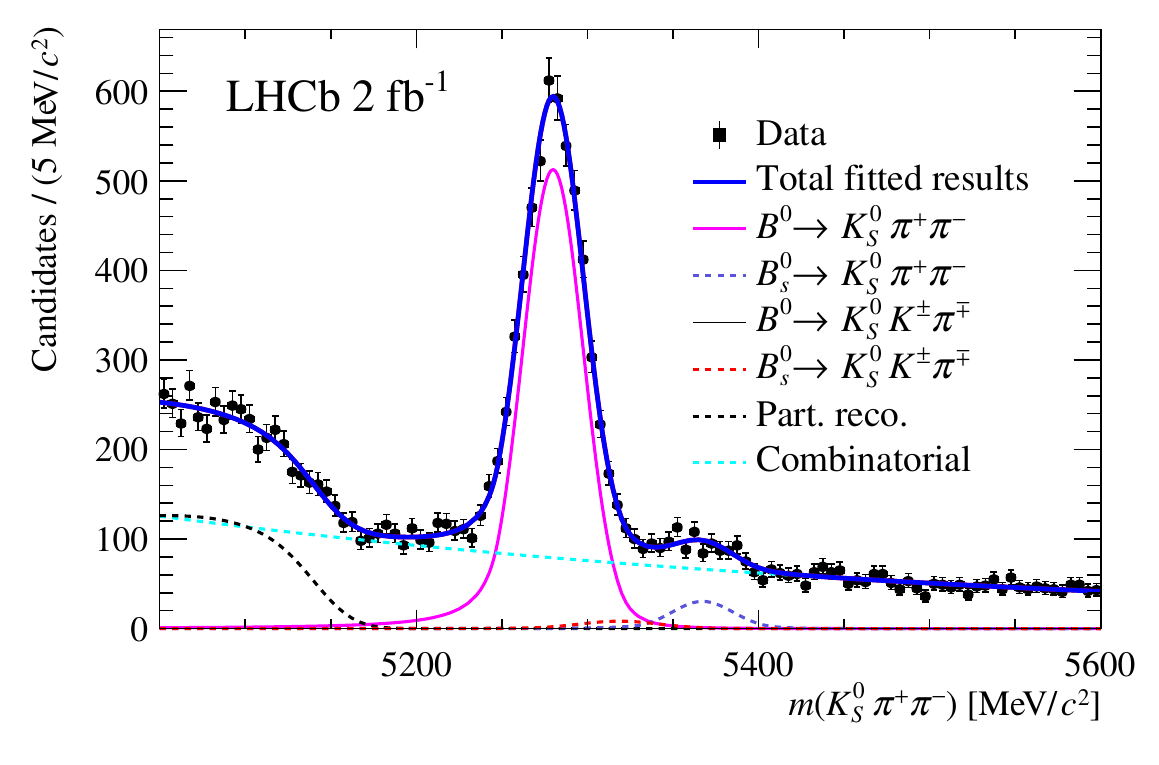}
    \includegraphics[width=0.45\textwidth]{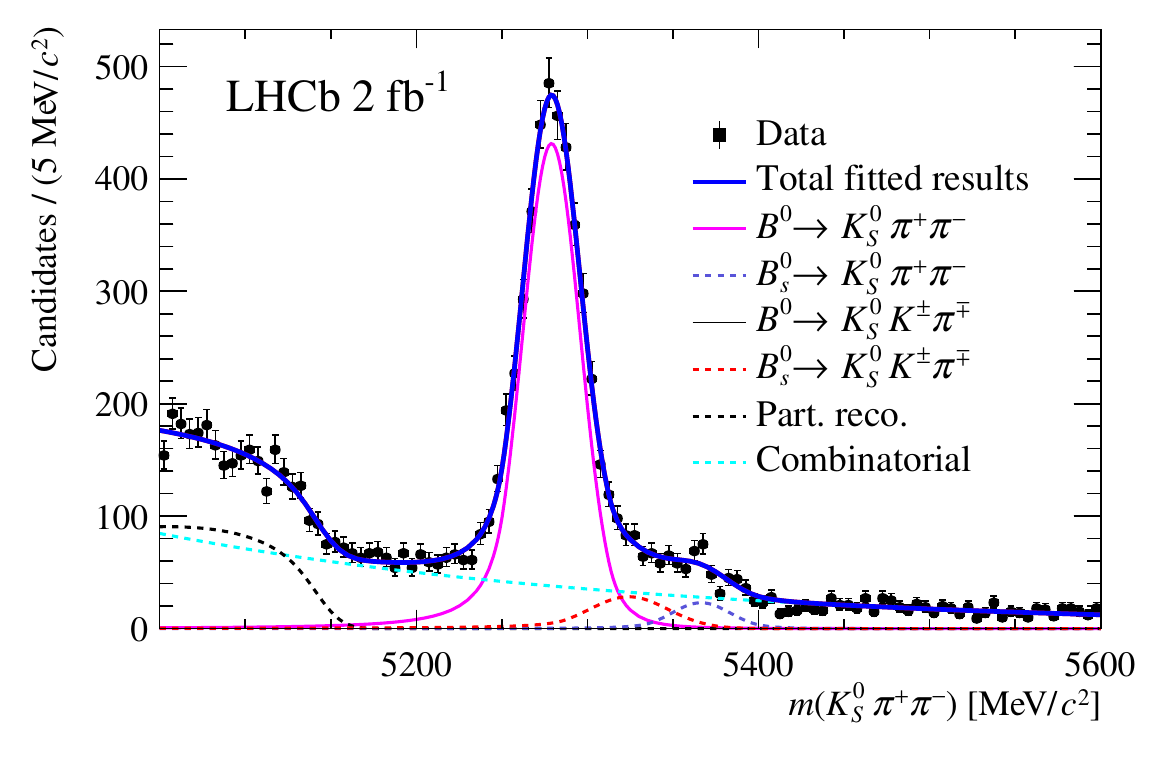}
    \includegraphics[width=0.45\textwidth]{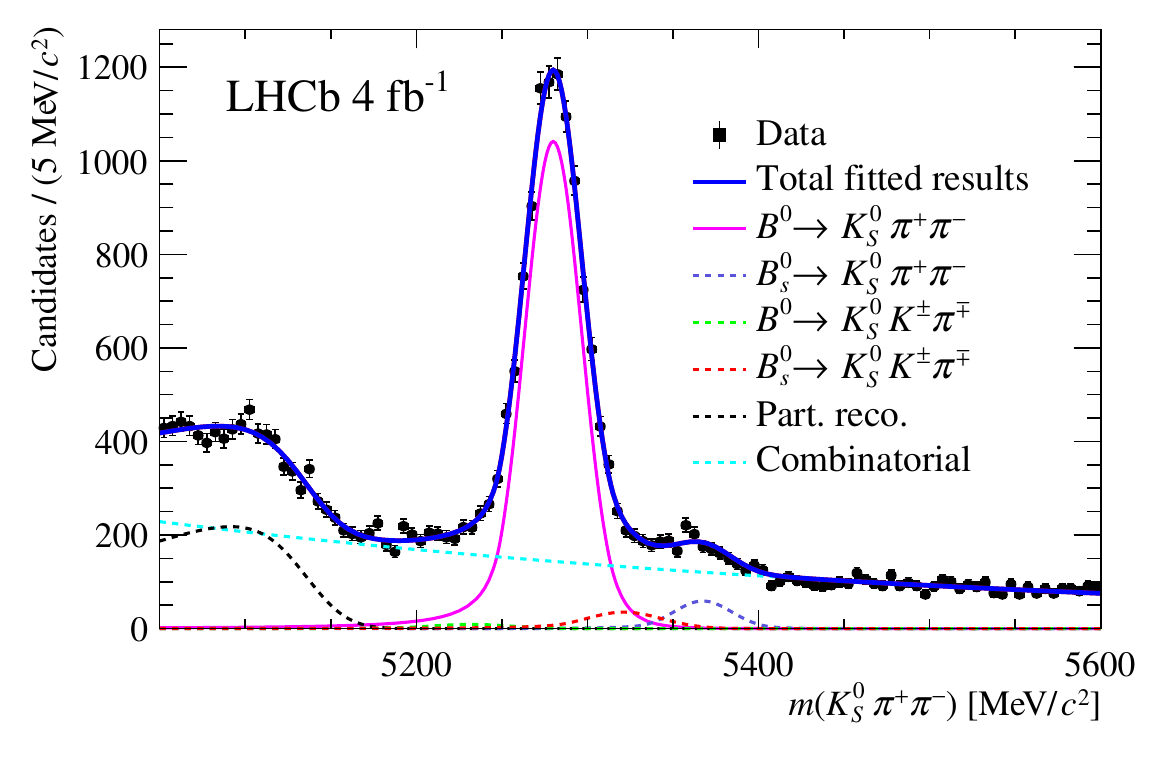}
    \includegraphics[width=0.45\textwidth]{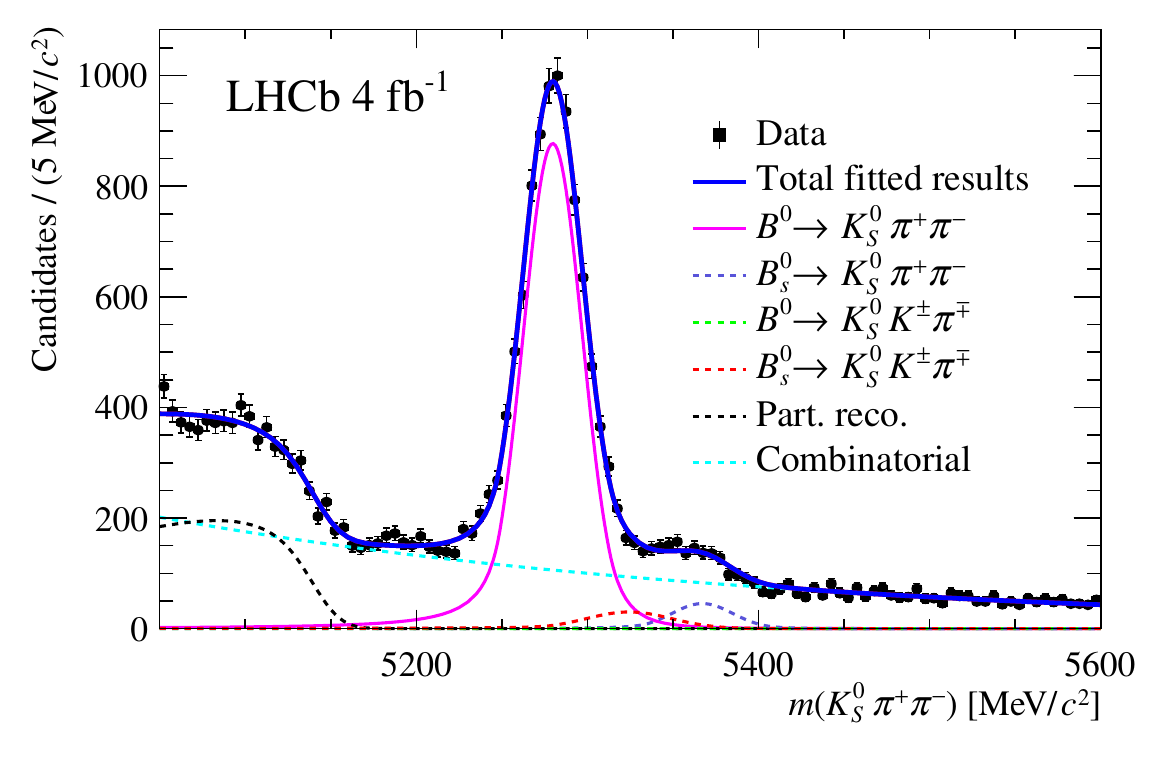}
  \end{center}
  \caption{Invariant-mass distributions of \KSpipi candidates passing the \BdToKSpipi selection. Data corresponding to (left) downstream and (right) long \KS reconstruction categories are shown for (top) the 2015--2016 dataset and (bottom) the 2017--2018 dataset. Fit results are also shown.}
  \label{Run2_BdKSpipi}
\end{figure}

\begin{figure}[htb]
  \begin{center}\includegraphics[width=0.45\textwidth]{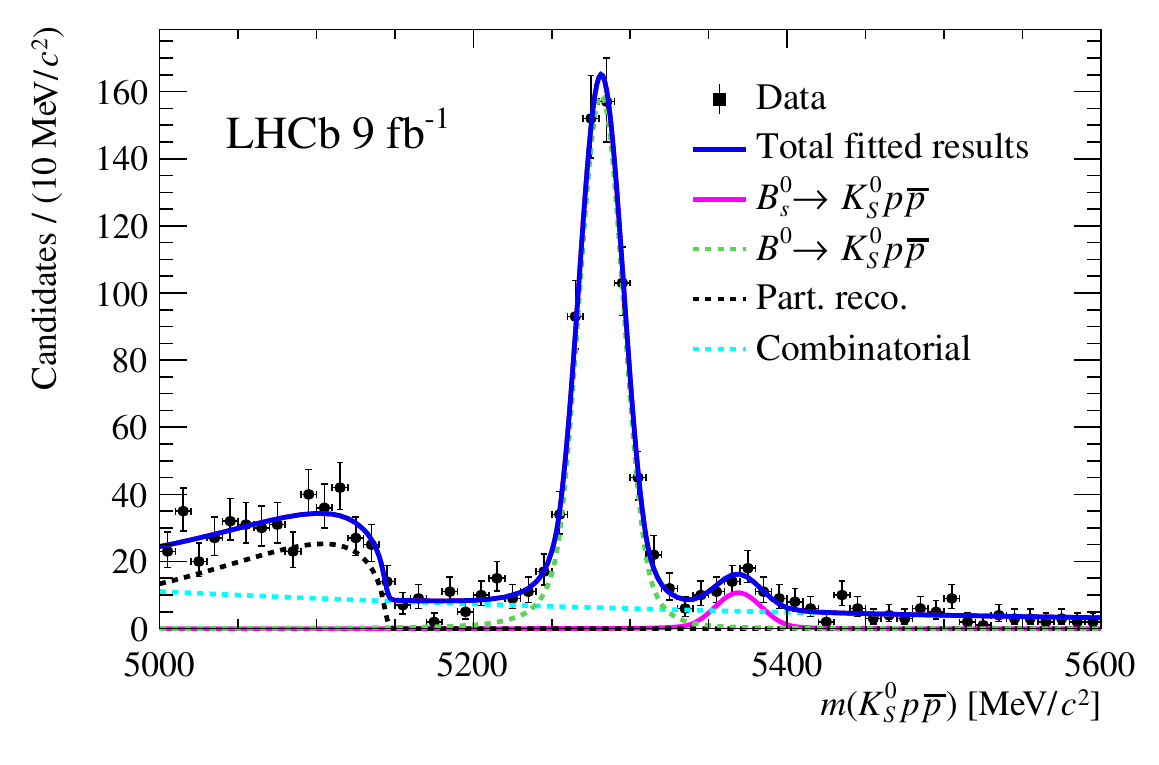}
    \includegraphics[width=0.45\textwidth]{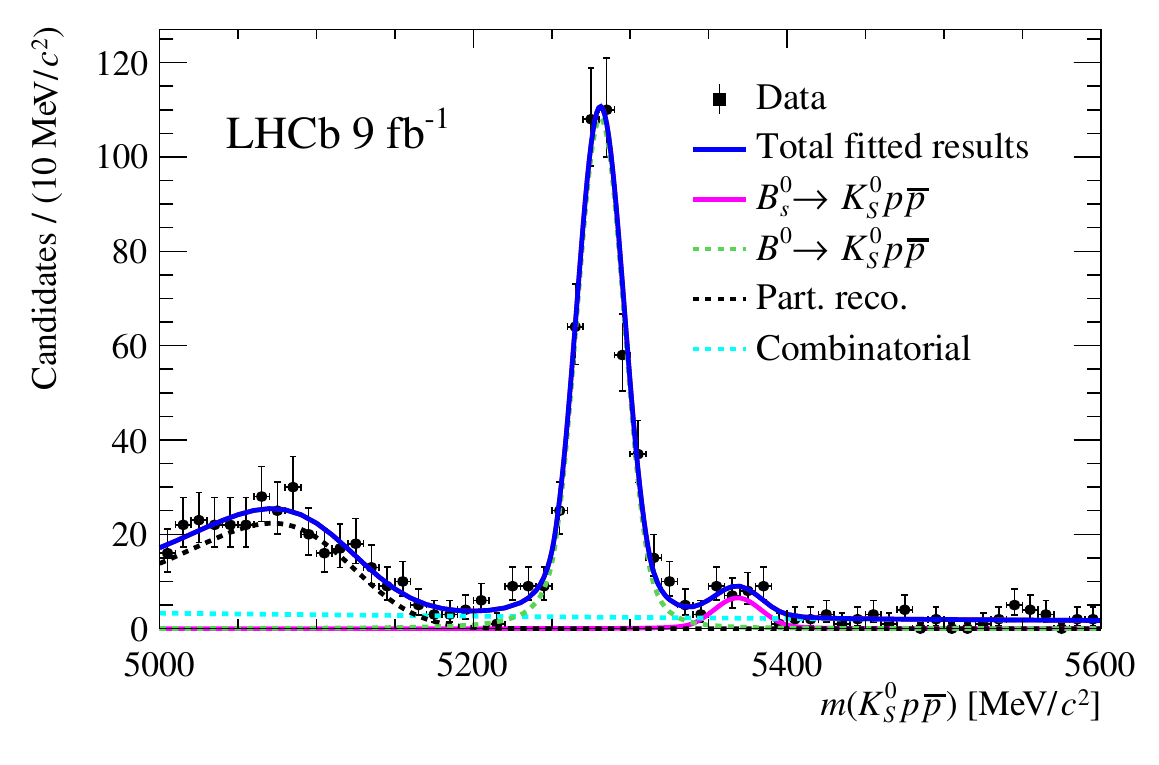}
  \end{center}
  \caption{Invariant-mass distributions of \KSpp candidates passing the \BsToKSpp selection in the full dataset for (left) downstream and (right) long \KS reconstruction categories. Fit results are also shown.}
  \label{fig:selection:fit_restlt_data:BsKSpp}
\end{figure}

The invariant-mass distributions for \BdToKSpp candidates (optimised for the branching fraction measurement) are shown in Figs.~\ref{Run1_BdKSpp} and \ref{Run2_BdKSpp} for Run 1 and Run 2, respectively, while Figs.~\ref{Run1_BdKSpipi} and \ref{Run2_BdKSpipi} present the corresponding distributions for \BdToKSpipi candidates. The \BsToKSpp candidates (optimised for search) distributions appear in Fig.~\ref{fig:selection:fit_restlt_data:BsKSpp}. All analyses distinguish between downstream and long \KS reconstruction types, with the corresponding fit results and signal yields summarised in Table~\ref{tab:fit_signal:yields}.

%%%%%%%%%%%%%%%%%%%%%%%%%%%%%%%%%%%%%%%%%%%%%%%%%%%%%%%%%%%%%%%%%%
%%%%%%%%%%%%%%%%%%%%%%%%%%%%%%%%%%%%%%%%%%%%%%%%%%%%%%%%%%%%%%%%%%

%%%%%%%%%%%%%%%%%%%%%%%%%%%%%%%%%%%%%%%%%%%%%%%%%%%%%%%%%%%%%%%%%%
%%%%%%%%%%%%%%%%%%%%%%%%%%%%%%%%%%%%%%%%%%%%%%%%%%%%%%%%%%%%%%%%%%

%%%%%%%%%%%%%%%%%%%%%%%%%%%%%%%%%%%%%%%%%%%%%%%%%%%%%%%%%%%%%%%%%%
%%%%%%%%%%%%%%%%%%%%%%%%%%%%%%%%%%%%%%%%%%%%%%%%%%%%%%%%%%%%%%%%%%

%%%%%%%%%%%%%%%%%%%%%%%%%%%%%%%%%%%%%%%%%%%%%%%%%%%%%%%%%%%%%%%%%%
%%%%%%%%%%%%%%%%%%%%%%%%%%%%%%%%%%%%%%%%%%%%%%%%%%%%%%%%%%%%%%%%%%

%%%%%%%%%%%%%%%%%%%%%%%%%%%%%%%%%%%%%%%%%%%%%%%%%%%%%%%%%%%%%%%%%%
%%%%%%%%%%%%%%%%%%%%%%%%%%%%%%%%%%
%%%%%%%%%%%%%%%%%%%%%%%%%%%%%%%%%%%%%%%%%%%%%%%%%%%%%%%%%%%%%%%%%%
\begin{table}[H]
    \caption{Signal yields for different decay channels and subsamples.
The yields for the \BdToKSpp decay are obtained from fits shown in  Figs.~\ref{Run1_BdKSpp} and \ref{Run2_BdKSpp}, while the yields for the \BsToKSpp decay correspond to fits shown in Fig.~\ref{fig:selection:fit_restlt_data:BsKSpp}. The quoted uncertainties are statistical only.}
    \centering
    {\fontsize{10}{12}\selectfont
    \begin{tabular}{c|c|
        r@{\,±\,}l
        r@{\,±\,}l
        r@{\,±\,}l
        r@{\,±\,}l}
    \hline
    \hline
    Modes & \KS & \multicolumn{2}{c}{2011} & \multicolumn{2}{c}{2012} & \multicolumn{2}{c}{2015--2016} & \multicolumn{2}{c}{2017--2018} \\ \hline
    \multirow{2}{*}{\BdToKSpp} & Downstream & 55 & 9 & 135 & 18 & 303 & 24 & 670 & 33 \\
                               & Long       & 25 & 7 & 83  & 12 & 174 & 18 & 346 & 25 \\ \hline
    \multirow{2}{*}{\BdToKSpipi} & Downstream & 1044 & 65 & 2653 & 96 & 4795 & 108 & 9203 & 143 \\
                                 & Long       & 772  & 52 & 1678 & 73 & 3892 & 85  & 8109 & 132 \\ \hline
    \multirow{2}{*}{\BsToKSpp} & Downstream & \multicolumn{8}{c}{41 $\pm$ 9} \\
                               & Long       & \multicolumn{8}{c}{26 $\pm$ 7} \\ \hline
    \hline
    \end{tabular}
    }
    \label{tab:fit_signal:yields}
\end{table}

\clearpage

\addcontentsline{toc}{section}{References}
%\setboolean{inbibliography}{true}
\bibliographystyle{LHCb}
\bibliography{main,standard,LHCb-PAPER,LHCb-CONF,LHCb-DP,LHCb-TDR}

\newpage
% LHCb collaboration author list
% Data extracted on April 18th, 2025 at 3:10am for paper reference LHCb-PAPER-2025-001
\centerline
{\large\bf LHCb collaboration}
\begin
{flushleft}
\small
R.~Aaij$^{38}$\lhcborcid{0000-0003-0533-1952},
A.S.W.~Abdelmotteleb$^{57}$\lhcborcid{0000-0001-7905-0542},
C.~Abellan~Beteta$^{51}$\lhcborcid{0009-0009-0869-6798},
F.~Abudin{\'e}n$^{57}$\lhcborcid{0000-0002-6737-3528},
T.~Ackernley$^{61}$\lhcborcid{0000-0002-5951-3498},
A. A. ~Adefisoye$^{69}$\lhcborcid{0000-0003-2448-1550},
B.~Adeva$^{47}$\lhcborcid{0000-0001-9756-3712},
M.~Adinolfi$^{55}$\lhcborcid{0000-0002-1326-1264},
P.~Adlarson$^{84}$\lhcborcid{0000-0001-6280-3851},
C.~Agapopoulou$^{14}$\lhcborcid{0000-0002-2368-0147}
% ...

{\footnotesize \it

% ...
$^{14}$Universit{\'e} Paris-Saclay, CNRS/IN2P3, IJCLab, Orsay, France\\
% ...
$^{38}$Nikhef National Institute for Subatomic Physics, Amsterdam, Netherlands\\
% ...
$^{47}$Instituto Galego de F{\'\i}sica de Altas Enerx{\'\i}as (IGFAE), Universidade de Santiago de Compostela, Santiago de Compostela, Spain\\
% ...
$^{51}$Physik-Institut, Universit{\"a}t Z{\"u}rich, Z{\"u}rich, Switzerland\\
% ...
$^{55}$H.H. Wills Physics Laboratory, University of Bristol, Bristol, United Kingdom\\
% ...
$^{57}$Department of Physics, University of Warwick, Coventry, United Kingdom\\
% ...
$^{60}$School of Physics and Astronomy, University of Glasgow, Glasgow, United Kingdom\\
$^{61}$Oliver Lodge Laboratory, University of Liverpool, Liverpool, United Kingdom\\
% ...
$^{69}$Syracuse University, Syracuse, NY, United States\\
% ...
$^{84}$Department of Physics and Astronomy, Uppsala University, Uppsala, Sweden, associated to $^{60}$\\
% ...
}
\end{flushleft}

\end{document}